\documentclass[twocolumn, a4paper, superscriptaddress, pra]{revtex4-1}

\usepackage{graphicx, epsfig, fancybox}
\usepackage{amssymb,amsmath}
\usepackage{color}

\usepackage{times}

\def \hf{\tfrac{1}{2}}    

\def \ord{\mathcal{O}}

\newcommand{\bra}[1]{\langle\left.{#1}\right|}
\newcommand{\ket}[1]{\left|{#1}\right.\rangle}
\newcommand{\abs}[1]{\left|{#1}\right|}
\newcommand{\braket}[2]{\langle{#1}\left.\right| {#2}\rangle}

\newcommand{\xpct}[1]{\langle{#1}\rangle}    % expectatn value

\begin{document}

\title{Quantum quenches and work distributions in ultra-low-density systems}

%\title{Quantum quenches in a system with competing itinerancy and binding effects, and overlap
% distributions in ultralow-density systems}

% \title{Quenches in a ??? model}

%\author{Y S}
\author{Yulia E. Shchadilova}
 \affiliation{Max Planck Institute for the Physics of Complex
 Systems, N\"othnitzer Str.~38, 01187 Dresden, Germany}

\affiliation{A. M. Prokhorov General Physical Institute, 
Russian Academy of Sciences, Vavilova str.~38, 119991 Moscow, Russia}

%\author{P R}
\author{Pedro Ribeiro}

%\author{M H}
\author{Masudul Haque}

 \affiliation{Max Planck Institute for the Physics of Complex
 Systems, N\"othnitzer Str.~38, 01187 Dresden, Germany}

\begin{abstract}

We present results on quantum quenches in systems with a fixed number of particles in a large
region.  
We show that the typical differences between local and global quenches present in systems with
regular thermodynamic limit are lacking in this low-density limit.
In particular, we show that in this limit local quenches may not lead to equilibration to the new
ground state, and that global quenches can have power-law work distributions (``edge
singularities'') typically associated with local quenches for finite-density systems.  We also show
that this regime allows for large edge singularity exponents beyond that allowed by the constraints
of the usual thermodynamic limit.  This large-exponent singularity has observable consequences in
the time evolution, leading to a distinct intermediate power-law regime in time.
We demonstrate these results first using local quantum quenches in a low-density Kondo-like system,
and additionally through global and local quenches in Bose-Hubbard, Aubry-Andre, and hard-core boson
systems in the low-density regime.

%% We demonstrate and illustrate these results using local quantum quenches in a low-density model
%% constructed to display the interplay of binding, boundary effects, itinerancy, and impurity (Kondo)
%% physics.  This is supplemented by results on quantum quenches in Bose-Hubbard, Aubry-Andre, and
%% hard-core boson systems in the low-density regime.

\end{abstract}

%\pacs{???, ???, ???}

%% 67.85.-d   Ultracold gases, trapped gases
%% 67.85.De   Dynamic properties of condensates; excitations, and superfluid flow 
%% 67.85.Jk   Other Bose-Einstein condensation phenomena
%%
%% 03.75.-b   Matter waves 
%% 03.75.Kk   Dynamic properties of condensates; collective and hydrodynamic excitations, superfluid flow
%% 03.75.Nt   Other Bose-Einstein condensation phenomena
%%
%% 67.10.-j   Quantum fluids: general properties
%% 67.10.Fj   Quantum statistical theory
%%
%% 05.70.-a   Thermodynamics 
%% 05.70.Ln   Nonequilibrium and irreversible thermodynamics
%%
%% 75.10.Pq   Spin chain models
%% 03.65.Xp   Tunneling, traversal time, quantum Zeno dynamics
%% 03.67.Hk   Quantum communication
%% 75.10.Jm   Quantized spin models
%% 75.45.+j   Macroscopic quantum phenomena in magnetic systems

\maketitle

\paragraph*{Introduction.}

Motivated by remarkable experimental progress in realizing and exploring non-equilibrium physics in
cold-atom systems \cite{cold_atoms_expts}, there has been increasing interest in the dynamics of
thermally isolated systems \cite{noneq_reviews}.  
%
% In isolated-system experiments, importance of boundaries and finiteness (cite Roux). 
%
Despite the rapidly growing body of research in this class of non-equilibrium dynamics, many aspects
are still poorly understood.  For example, what type of equilibration can be expected for various
types of local and global quenches?  
Another question involves the \emph{overlap distribution}, closely related to the \emph{work
  distribution} \cite{work_distribution,Silva_PRL2008, work_dist_condmat_models} for a quantum
quench.  What is the typical form of the distribution of overlaps of the initial state with the
final eigenstates?  What are the effects of various overlap distributions and work distributions on
dynamical (time-evolving) quantities?

In the experimental settings suitable for exploring non-equilibrium physics, such as cold atoms and
semiconductor nanostructures like quantum wells, a common situation is to have a fixed number of
particles in a large spatial region.  This contrasts sharply with the solid-state notion of the
thermodynamic limit, where large regions are filled with a constant density.  The study of
non-equilibrium issues (e.g., quenches and work distributions) in such situations, where the usual
thermodynamic limit is not applicable, is clearly of topical importance but has been near-absent in
the non-equilibrium theory literature.

In this work, we focus on this ultra-low-density limit --- fixed number of particles, arbitrary large
sizes.  We present a study of quenches in a system which is the counterpart of the Kondo model in
this low-density regime.  We present several dynamical aspects which, through calculations in a few
other low-density systems, we show to be generic features of quantum quenches in this limit.

One peculiarity of this limit is a blurring of differences between the consequences of local versus
global quenches.
Another striking result involves the overlap distribution,
$\left|\langle\Psi(0)\left|\right.\phi_m^{(f)}\rangle\right|$, where
$\ket{\Psi(0)}=\ket{\phi_0^{(i)}}$ is the initial state (ground state of initial Hamiltonian), and
$m$ indexes the eigenstates of the final Hamiltonian.  We show that this quantity is dominated by a
power-law decay, ${\sim}m^{-\alpha}$, generically for quenches involving low-density systems.  The
associated ``edge singularity'' in the work distribution has large power-law exponents which would
not be compatible with the usual thermodynamic limit.  This in turn has remarkable consequences on
the real-time evolution: in the evolution of observables away from their initial value, there
appears an intermediate power-law regime between the initial perturbative time period and the
large-time steady-state behavior.

%
%% In systems with constant finite density, a local quench is expected to be followed by equilibration
%% to the new ground state \emph{sans} $1/L$ corrections.  In a system with constant particle number, a
%% local quench may well lead to equilibration at values quite different from the new ground state, as
%% we will see through examples.

Since the fixed-number large-size limit is applicable to many experimental non-equilibrium setups,
these results are expected to be relevant to experimental situations realized or realizable in the
near future.

\paragraph*{Kondo-like model.}

\begin{figure}[t]
\includegraphics[width=1\columnwidth]{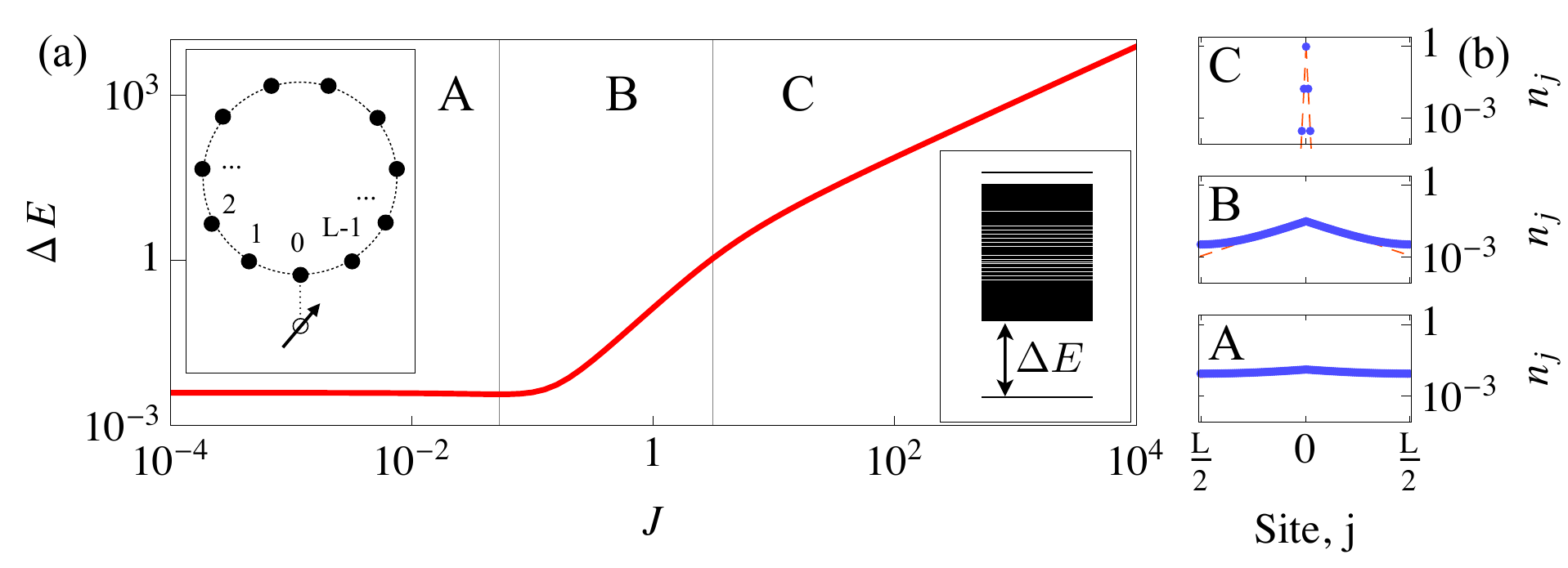}
\caption{ \label{fig:intro}
Kondo-like model with single mobile fermion.  
(a) Energy gap as a function of $J$, $L=100$ sites, showing distinct behaviors in three regimes.
Boundaries between A,B (B,C) regimes are denoted according to criteria $\xi=L$ ($\xi=1$). ($\xi=$
localization length.)  Left inset shows system geometry.  Right inset shows spectrum in C regime.
(b) Density profiles $n_j$ in three regimes; dashed lines showing exponential localization.
}
\end{figure}

The main system we use for demonstrating these general results involves a few ($N_c$) mobile
fermions (``conduction electrons'') in a tight-binding closed chain (Fig.\ \ref{fig:intro}).  One
site of the lattice is Kondo-coupled to a single spin-$\hf$ ``impurity''.  The Hamiltonian is
\begin{equation} \label{eq:minimalKondoHam}
H ~=~ - \sum_{i,s} \left( c_{i,s}^{\dagger}c_{i+1,s} +
% c_{i+1,s}^{\dagger}c_{i,s}\rba
\mathrm{h.c.} \right)
 ~+~  J  \vec{S}_{\mathrm{imp}}\cdot \vec{\mathcal{S}_0}
\end{equation}
where $\vec{\mathcal{S}_0}=\sum_{s,s'}c_{0,s}^{\dagger}\vec{\sigma}_{ss'}c_{0,s'}$ is the spin on
site $i=0$ ($s$, $s'$ are spin indices), and $i\in[0,L-1]$ is the site index.  We study quenches of
$J$, i.e., local quenches, starting from the ground state at $J=J_i$ and studying the dynamics after
changing $J$ instantaneously to its new value $J_f$.  The ground state is a spin singlet, and
quenches of $J$ preserve the spin, so that all dynamics is confined to the spin singlet sector.
%
%% The states are therefore primarily characterized by the density profile of the itinerant fermions.
%% The Kondo coupling serves as a potential that tries to localize the itinerant particles.

In Fig.\ \ref{fig:intro} we summarize the equilibrium physics of the $N_c=1$ system.  In an infinite
chain, in the ground state, the fermion is localized around the impurity-coupled site ($i=0$) with
localization length $\xi$.  ($\xi$ decreases with increasing $J$.)  At large $J$ (regime C), the
itinerant fermion is almost completely localized at site 0 ($\xi\lesssim1$).  At smaller $J$, the
itinerant fermion is spread over multiple sites $\xi>1$ (regime B).  In an infinite system, this
region would extend to arbitrarily small $J$.  However, for any finite size $L$, there is a
boundary-sensitive small-$J$ regime (regime A) where the fermion cloud extends over the whole system
($\xi\gtrsim{L}$).
For $1<N_c{\ll}L$, we naturally get additional features, but the same general behavior persists in
the three regimes.

\paragraph*{Observables.}

We will present time dependences of the occupancy $n_0(t)$ of site $i=0$ for the Kondo-like system,
and of the Loschmidt echo $\mathcal{L}(t) = \left|\langle\Psi(0)\left|\Psi(t)
\right.\rangle\right|^2$.  The observable $n_0(t)$ is of obvious importance for the model
\eqref{eq:minimalKondoHam}, while $\mathcal{L}(t)$ is well-defined for any model and is closely
related to the work distribution \cite{work_distribution,Silva_PRL2008, work_dist_condmat_models}.
Despite the nonlocal nature of the Loschmidt echo, there exist proposals for experimentally
measuring this quantity, and related quantities have been measured \cite{LE_measurement}.

\paragraph*{Lack of equilibration to new ground state in local quenches.}

\begin{figure}[t]
\includegraphics[width=1\columnwidth]{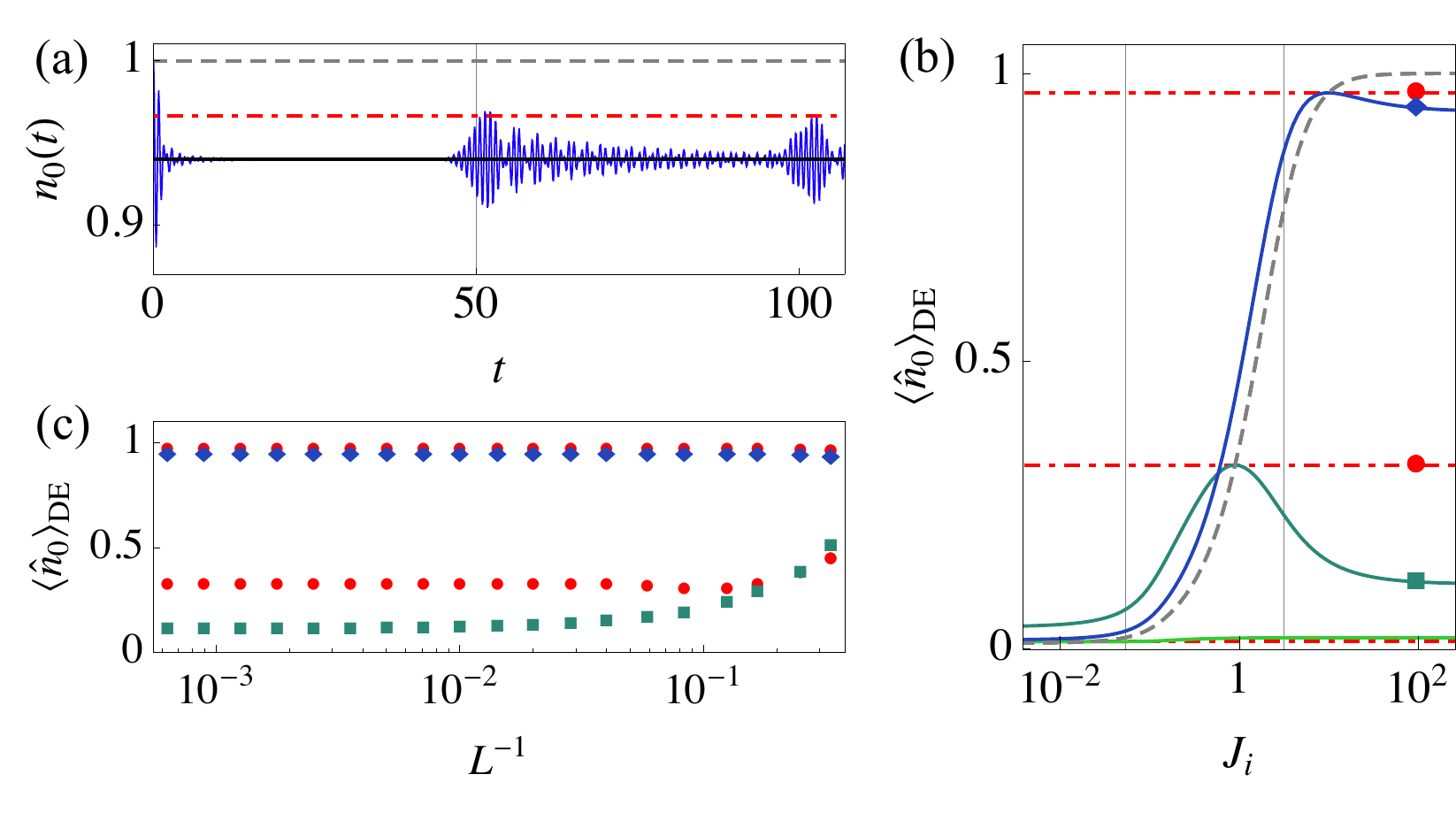}
\caption{ \label{fig:DiagEnsemb} 
Local quench in Kondo-like model with $N_c=1$ mobile fermion. 
(a) Time evolution of occupancy $n_0$ at impurity-coupled site; C$\to$C quench, $J_i=10^{2}$,
  $J_f=10$; $L=100$ sites.  The long-time average $\xpct{\hat{n}_0}_{DE}$ (solid black line), around
  which $n_0(t)$ oscillates, is significantly different from the equilibrium $J=J_f$ value.
(b) $\xpct{\hat{n}_0}_{DE}$ as function of $J_i$, for final values in A,B,C regimes:
  $J_{f}=0.02$ (green), $0.9$ (dark green), $10$ (blue).  Dots mark the parameters for panel c. 
In (a,b), gray dashed and red dash-dotted lines show equilibrium $n_0$ values for
  $J=J_i$ and $J=J_f$.  
(c) Size dependence of $\xpct{\hat{n}_0}_{DE}$ values for $J_i=10^2$. Upper blue:  $J_f=10$;
lower green: $J_f=0.9$.  Red dots are corresponding $J=J_f$ equilibrium values.
}
\end{figure}

The final value at which an observable $\hat{O}$ saturates is given by
$\langle\hat{O}\rangle_{\mathrm{DE}} =
\sum_m\left|\langle\Psi(0)\left|\right.\phi_m^{(f)}\rangle\right|^2\langle\phi_m^{(f)}\left|\hat{O}\right|\phi_m^{f}\rangle$,
the so-called ``diagonal ensemble'' (DE) value \cite{RigolOlshanii_Nature2008}.
In Fig.\ \ref{fig:DiagEnsemb}(a) we show the time dependence of $n_0$ after a quench within the C
region.  
We note that $n_0(t)$ reaches the DE value $\xpct{\hat{n}_0}_{\mathrm{DE}}$ relatively rapidly, and
then shows `revivals' at roughly periodic intervals of $t\sim{L/2}$.  The DE value where $n_0$
saturates is markedly different from the ground state value of $n_0$ for $J=J_f$.  This seemingly
contradicts the intuition that a local quench in a large system should lead to relaxation to the
final ground state value, because the energy pumped into the system by a local quench is a
$\ord(L^{-1})$ effect.  The reason this does not happen in the C$\rightarrow$C quenches is that the
itinerant electron only occupies a small number of sites near the impurity position.  Thus, most of
the lattice sites do not play any role in the dynamics, and cannot serve as a bath to absorb the
disturbance at site 0.
This effect is not restricted to $N_c=1$, but is true for finite number $N_c>1$ of fermions for
$L\to\infty$ \cite{suppl}.  
%
%% The reasoning is that the sites away from the localization region have
%% vanishing $\ord(L^{-1})$ occupancy, and cannot absorb the effect of a local quench in the
%% localization region.

In Fig.\ \ref{fig:DiagEnsemb}(b) we show $\xpct{\hat{n}_0}_{\mathrm{DE}}$ as a function of $J_i$ for
fixed $J_f$.  The $\xpct{\hat{n}_0}_{\mathrm{DE}}$ values deviate significantly from the $J=J_f$
equilibrium values for most $J_i$, $J_f$ combinations.
Fig.\ \ref{fig:DiagEnsemb}(c) shows, through $L$-dependences of $\xpct{\hat{n}_0}_{\mathrm{DE}}$,
that the lack of equilibration in quenches to C or B regions is not a finite-$L$ effect.

This effect represents a loss of the distinction between local and global quenches, which is a
generic feature of the $L\to\infty$ limit with finite particle number.

\paragraph*{Overlap Distributions.}

\begin{figure}[t]
\includegraphics[width=1\columnwidth]{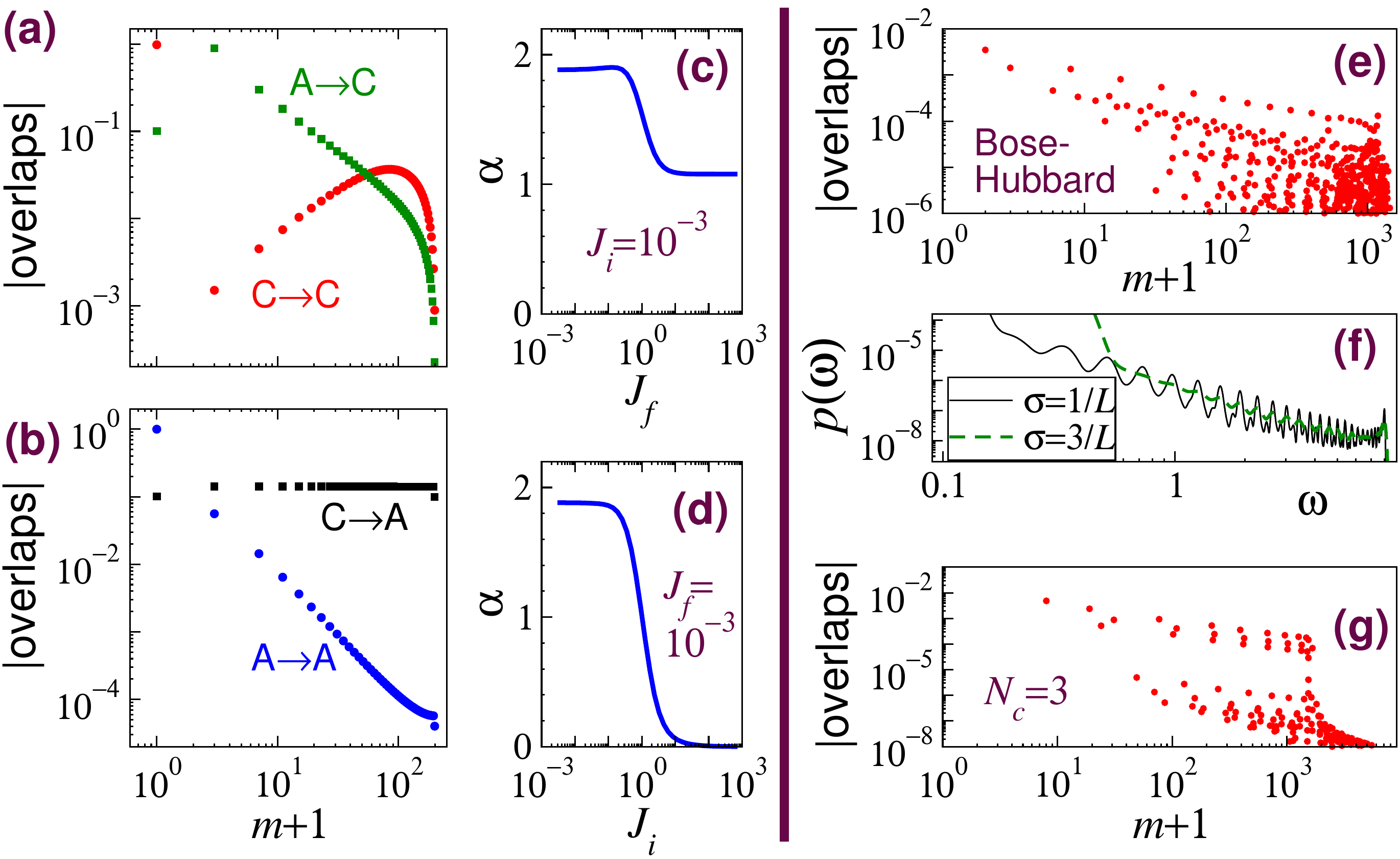}
\caption{  \label{fig:overlaps} 
Overlap distribution $\left|\langle
  \phi_0^{(i)}\left|\right.\phi_m^{(f)}\rangle\right|$, power-law exponents $\alpha$, and work
  distribution $p(\omega)$.    
(a-d) Kondo-like model, $N_c=1$.  (a,b) Quenches from $J_i=10^{-3}$ to
  $J_f=10^{3}$ (A$\to$C), $J_i=10$ to $J_f=10^{3}$ (C$\to$C), $J_i=10^{3}$ to $J_f=10^{-3}$
  (C$\to$A), and $J_i=10^{-3}$ to $J_f=10^{-2}$ (A$\to$A).  (c,d) Exponents of power-law fits
  $m^{-\alpha}$ for quenches ending at and starting from  the A regime.  (e,f) Bose-Hubbard chain
  with $N_b=3$ bosons in $L=20$ sites; interaction $U$ quenched from 0.3 to 0.5.  (f)
  Approximations to the work distribution are obtained using Gaussians of width $\sigma$ to replace
  the delta function of Eq.\ \eqref{eq:work_dist_def}.  (g)    Kondo-like model, $N_c=3$; quench
  from $J_i=10^{-3}$ to $J_f=10^{-1}$. 
}
\end{figure}

Fig.\ \ref{fig:overlaps}(a-d) summarize overlap distribution behaviors in  quantum quenches
between different regimes of the system \eqref{eq:minimalKondoHam} for $N_c=1$.  These behaviors can
be derived from detailed consideration of the eigenfunctions \cite{suppl}.
In C$\to$C quenches, the ground state overlap
$\left|\langle\phi_0^{(i)}\left|\right.\phi_{m=0}^{(f)}\rangle\right|$ is much larger than
the others, while the small $m\neq0$ overlaps have the form $\propto\sin(c_1{m})$ \cite{suppl}.
%
% obey $\sim\sin({\pi}m/2L)$ \cite{suppl}.
%
The most remarkable feature is the power-law behavior, $\left|\langle
\phi_0^{(i)}\left|\right.\phi_m^{(f)}\rangle\right| \sim m^{-\alpha}$, in quenches starting from or
ending in the A region.  The exponent $\alpha$ is 2 for A$\to$A quenches and 1 for A$\to$C quenches.

These power law behaviors are a generic phenomenon; we have found such power-law overlap
distributions in several other systems in the low-density limit, both for local and global quenches.
(The behavior is particularly clean for the $N_c=1$ system because of its simplicity.)
Fig.\ \ref{fig:overlaps}(g) shows the overlap distribution for the same model with $N_c=3$ fermions.
There are now additional structures, but the dominant overlaps follow a clear power law.
Fig.\ \ref{fig:overlaps}(e) shows the overlap distribution for a Bose-Hubbard chain at low density
\cite{suppl}.  Again, there are interesting additional structures, but the dominant overlaps follow a
clear power law ($\sim{}m^{-\alpha}$).

\paragraph*{Work distribution.}

The overlap distribution is related to 
\begin{equation}  \label{eq:work_dist_def}
p(\omega) = \sum_{m} \delta(\omega-\epsilon_m) \left|\langle \phi_0^{(i)}\left| \right. \phi_m^{(f)}
\rangle\right|^2 
\end{equation}
where $\epsilon_m=E_m^{(f)}-E_0^{(f)}$ are the final eigenenergies measured from the final ground
state energy.  This is the so-called work distribution \cite{work_distribution,Silva_PRL2008,
  work_dist_condmat_models}, except for a shift between $\omega$ and the usual work variable.  (The
energy prior to the quench plays no role in the temporal dynamics and so is not relevant for this
work.)  The work distribution is related to the Loschmidt echo: $\mathcal{L}(t) =
\left|\int{}d\omega{}p(\omega)e^{i{\omega}t}\right|^2$.  Since $\mathcal{L}(0)=1$ by definition,
$p(\omega)$ must be normalizable.

At large sizes (but constant particle number), $p(\omega)$ can be treated as a continuous function starting from
$\omega=\Delta$, the finite-size gap, which vanishes at large $L$.    
We have found that, for quantum quenches in low density systems, the work distribution generically
has behavior $p(\omega)\approx{}p_0\omega^{-b}$ for $\omega>\Delta$, with large exponents $b>1$.
These power-law divergences are analogs of what would be called ``X-ray edge singularities'' in
systems with a regular thermodynamic limit.  In finite-number systems, $p(\omega)$ remains
normalizable despite the singularity as $\Delta\to0$ because the magnitude of $p(\omega)$ also
vanishes ($p_0\to0$) in the large-size limit, due to the vanishing density.

This contrasts sharply to systems with the usual thermodynamic limit where density remains constant
as $L\to\infty$, and $p(\omega)$ itself is a well-defined non-vanishing quantity in the limit.  This
constrains the singularity $p(\omega)\sim\omega^{-b}$ to have smaller exponent, $b<1$ (e.g.,
\cite{Silva_PRL2008}).  The low-density systems of interest here have no such constraint; a central
result of the present work is that super-linear singularities ($b>1$) are signatures of low-density
systems.

%% For the model \eqref{eq:minimalKondoHam} with $N_c=1$, the behavior ${\sim}m^{-\alpha}$ implies
%% energy-dependence ${\sim}\omega^{-\alpha/2}$ for the overlap distribution.  Together with a factor of
%% $\omega^{-1/2}$ from the density of states, this leads to $p(\omega)\sim\omega^{-\alpha-1/2}$, i.e.,
%% $p(\omega)\sim\omega^{-5/2}$ for A$\to$A quenches and $p(\omega)\sim\omega^{-3/2}$ for A$\to$C
%% quenches ($b>1$ in both cases).

For the model \eqref{eq:minimalKondoHam} with $N_c=1$, $p(\omega)\sim\omega^{-5/2}$ (A$\to$A) and
$p(\omega)\sim\omega^{-3/2}$ (A$\to$C).  ($b>1$ in both cases.)
Fig.\ \ref{fig:overlaps}(f) shows the work distribution for a \emph{global} interaction quench in
the Bose-Hubbard chain, with the delta function regularized as gaussian.  There is a power law with
super-linear ($b>1$) singularity.  This is another example of the loss of distinction between global
and local quenches in the low-density limit, as ``edge singularities'' are normally associated only
with \emph{local} quenches for finite-density systems \cite{Silva_PRL2008}.

%% We will see later that ``edge singularities'',
%% normally associated with local quenches for large finite-density systems \cite{Silva_PRL2008},
%% appear for both local and global quenches for large-size finite-number systems.

We have also found super-linear singularity exponents in other low-density systems
\cite{suppl}, e.g., quenches of the strength/position of a weak trapping potential for a
Bose-Hubbard system, quenches of on-site potentials and hopping strengths for hard-core bosons in a
ladder geometry, and quenches of quasi-disorder potential strengths in an Aubry-Andr\'e
\cite{AubryAndre} system.

\paragraph*{Role of the density of states.}

For the model \eqref{eq:minimalKondoHam} with $N_c=1$, the behavior ${\sim}m^{-\alpha}$ implies
energy-dependence ${\sim}\omega^{-\alpha/2}$ for the overlap distribution.  Together with a factor
of $\omega^{-1/2}$ from the 1D single-particle density of states, this leads to
$p(\omega)\sim\omega^{-\alpha-1/2}$, i.e., 
$b=5/2 (3/2)$ for A$\to$A(C) quenches.
%
%% $p(\omega)\sim\omega^{-5/2}$ for A$\to$A quenches and $p(\omega)\sim\omega^{-3/2}$ for A$\to$C
%% quenches.

This argument can be generalized: if the overlap distribution follows $m^{-\alpha}$ and the density
of states in the relevant lower-energy part of the spectrum behaves as $\rho(\omega)\sim
\omega^{\gamma}$, the work distribution $p(\omega)\sim\omega^{-b}$ will have exponent
$b=2\gamma\alpha-\gamma+2\alpha$ \cite{suppl}.  For single-particle systems,we have $\gamma=-1/2$ in
1D, as in the above example.   
For a generic system, however, the \emph{many-body} density of states does not necessarily behave as
a power law.  We have found cases (Bose-Hubbard chain with trap) where an approximate power-law
region with exponent $\tilde{\gamma}$ in $\rho(\omega)$ leads to an approximate power law in
$p(\omega)$ with exponent $\tilde{b}= 2\tilde{\gamma}\alpha-\tilde{\gamma}+2\alpha$ \cite{suppl}.
Also, if $\alpha=1/2$, \emph{any} power-law form of $\rho(\omega)$ implies a linear edge singularity
$\rho(\omega)\sim\omega^{-1}$.  In this case, a super-linear edge singularity can only happen with
some non-power-law form of $\rho(\omega)$.  This occurs in the Bose-Hubbard chain case of
Fig.\ \ref{fig:overlaps}(e,f) \cite{suppl}.

\paragraph*{The intermediate-time $\sim{}t^{\beta}$ region.}  

\begin{figure}[t]
\includegraphics[width=1\columnwidth]{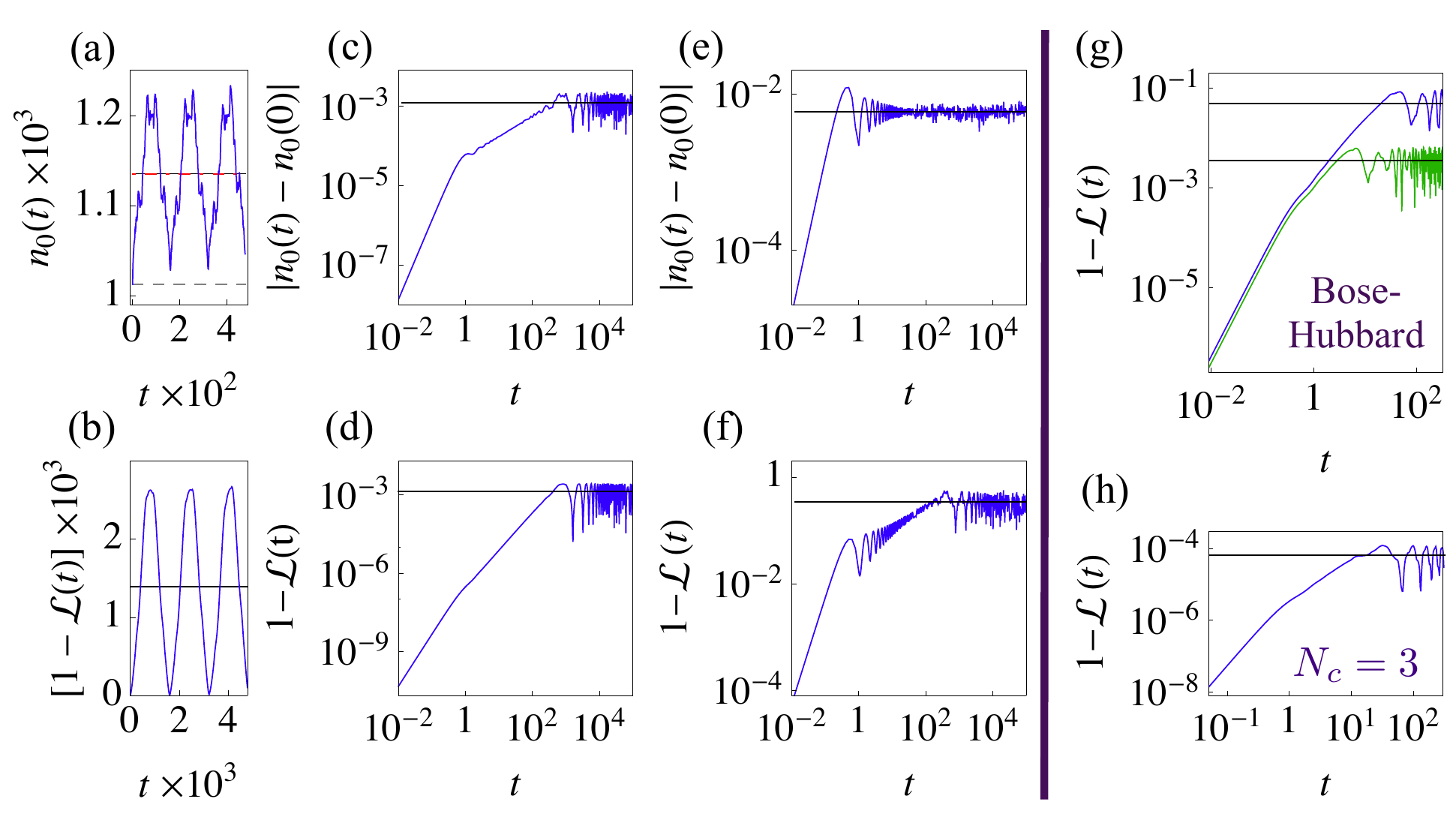}
\caption{  \label{fig:intermediate_powerlaw}
(a-f) Time evolution of  $n_0(t)$ and  $\mathcal{L}(t)$ in $N_c=1$ model.
(a-d) A$\to$A quenches: $J_i=10^{-3}$, $J_f=10^{-2}$.  (e.f) A$\to$C: $J_i=10^{-3}$, $J_f=10$.  
Extended intermediate region between ${\sim}t^2$ region and long-time oscillatory region is seen in
$n_0(t)$ for A$\rightarrow$A quenches and in $\mathcal{L}(t)$ for both A$\rightarrow$A and
A$\rightarrow$C quenches, but not in $n_0(t)$ for A$\rightarrow$C quenches.
(g) Bose-Hubbard, global interaction quench, 3 bosons in chains of length $L=10$ and $L=30$.  Quench
from $U_i=0.02L$ to $U_f=0.01L$.
(f)  $N_c=3$ fermions;  $J_i=10^{-3}$, $J_f=10^{-2}$.  
}
\end{figure}

The appearance of larger powers in the edge singularity has novel consequences for real-time
dynamics.  We have identified an intermediate-time power-law region in the dynamics of the Loschmidt
echo (and other observables), that appears as a direct consequence of the large-power edge
singularity.

At initial times after a quench, observables and $\mathcal{L}(t)$ evolve away from their initial
value quadratically with time, $\sim{t^2}$, as can be explained from generic perturbative arguments.
We have found that, when $p(\omega)$ has a large-exponent singularity, there is a region of time
(after the initial perturbative times and before the large-time steady-state oscillations), where
$\mathcal{L}(0)-\mathcal{L}(t) = 1-\mathcal{L}(t)$ follows a new power-law behavior.  If
$p(\omega)\sim\omega^{-b}$ with $b\in(1,3)$ in the energy range $[\Delta,\Lambda]$ and the
contributions outside this energy window can be neglected, then in the time window between
$t\sim\Lambda^{-1}$ and $t\sim\Delta^{-1}$ one sees the behavior $1-\mathcal{L}(t) \sim t^{b-1}$
\cite{suppl}.

The same phenomenon is also found in some observables: $|O(t)-O(0)|$ can have an extended region
after the initial perturbative $\sim{t^2}$ region with a new exponent $\sim{t^{\beta_{\hat{O}}}}$.
When $\hat{O}$ has the form of a rank-1 projector, $\hat{O}=\ket{\chi}\bra{\chi}$, we can write the
time evolution as $O(t) = \left|\int{}d\omega{}p_{\hat{O}}(\omega)e^{i{\omega}t}\right|^2$, where 
\begin{equation} \label{eq:p_O}
p_{\hat{O}}(\omega) = \sum_{m} \delta(\omega-\epsilon_m) \langle \phi_0^{(i)}\left| \right. \phi_m^{(f)}
\rangle \langle \phi_m^{(f)} \left| \right. \chi \rangle 
\end{equation}
differs from the work distribution \eqref{eq:work_dist_def} in that one factor of the overlap is
replaced by $\langle\phi_m^{(f)}\left|\right.\chi\rangle$.
If $p_{\hat{O}}(\omega)$ has a power-law singularity structure $\omega^{-b_{\hat{O}}}$ with exponent
$b_{\hat{O}}\in(1,3)$, the time evolution of $O(t)$ away from $O(0)$ will show the intermediate-time
region $\sim{}t^{\beta_{\hat{O}}}$ ($\beta_{\hat{O}}={b_{\hat{O}}-1}$).   
When the operator $\hat{O}$ does not have the form $\hat{O}=\ket{\chi}\bra{\chi}$, it is not simple
to formulate an analogous expression.  A generic operator for a many-body system will not have this
form, but the site occupancies for systems with single itinerant particles (e.g., $n_j$ for our
$N_c=1$ system) have the forms of rank-1 projectors, as does the Loschmidt echo for any system.
%
% , which in operator form is simply the system density matrix $\ket{\psi}\bra{\psi}$.
%
Currently, little is  known about $p_{\hat{O}}(\omega)$ behaviors for projector-type
observables in different quantum quenches, or about the conditions necessary for having an
intermediate-time regime in generic observables.

The intermediate-time regime for low-density systems is illustrated in
Fig.\ \ref{fig:intermediate_powerlaw}.  For the model \eqref{eq:minimalKondoHam} with $N_c=1$
particle, this regime is present in the Loschmidt echo, for both A$\to$A and A$\to$C quenches.  In
the occupancy $n_0(t)$, the intermediate-exponent regime can be seen for A$\to$A quenches (with form
${\sim}t^{1/2}$), but not for the A$\to$C quenches, for which case the eigenstate dependence of
$\langle\phi_m^{(f)}\left|\right.\chi\rangle$ does not favor a large enough exponent in
$p_{\hat{n}_0}(\omega)$ \cite{suppl}.
Fig.\ \ref{fig:intermediate_powerlaw}(g) displays the intermediate-time region for a Bose-Hubbard
chain with interaction quenches, and Fig.\ \ref{fig:intermediate_powerlaw}(h) shows the same for the
Kondo-like model with $N_c=3$ itinerant fermions.

With hard-core bosons on a ladder-shaped lattice, considering time evolution after various local and
global quenches of hopping strengths and on-site potentials, we find extended intermediate-time
regions $\sim{t^{\beta}}$ in $1-\mathcal{L}(t)$, with exponents matching $\beta=b-1$ where $b$ is
the singularity exponent in $p(\omega)$, calculated with gaussian regularization \cite{suppl}.  With
quenches of a trapping potential, we find quench parameter combinations where $p(\omega)$ shows
super-linear edge singularities ($b>1$) but no intermediate-time regime shows up in the
$\mathcal{L}(t)$ dynamics because the singularity exponents are \emph{too large}, $b>3$
\cite{suppl}.  We have also found an example (Aubry-Andr\'e system) where there are well-defined
$p(\omega)\sim\omega^{-b}$ regions but the contributions from outside the power-law region are so
large that the dynamical intermediate-time signature is washed out \cite{suppl}.

\paragraph*{Extent of the intermediate-time region.}  

If the power-law window for $p(\omega)$ is $\omega\in[\Delta,\Lambda]$, the ${\sim}t^{\beta}$ region
with $\beta\in(0,2)$ extends from $t\sim\Lambda^{-1}$ to $t\sim\Delta^{-1}$.  The scale $\Lambda$ is
generally of the order of the bandwidth, and so is set by the hopping strength.  Since the
finite-size gap $\Delta$ vanishes with increasing system size $L$, the intermediate-time region gets
more and more extended in time for larger $L$.  This is shown in
Fig.\ \ref{fig:intermediate_powerlaw}(g) through a comparison of two different $L$ values.

\paragraph*{Discussion.}

For systems that are not well-described by the traditional thermodynamic limit but instead have a
fixed number of particles in a large size, as is common in setups relevant for large classes of
non-equilibrium experiments, we have presented universal features of quantum quenches.  These
include edge singularities with large exponents not possible in `regular-limit' systems, a loss of
the usual distinctions between local and global quenches, and a novel intermediate-time region in
the dynamics.

Universal behaviors in quantum quenches are generally sought and discussed in asymptotic times.  A
new universality at intermediate times, visible in widely different systems, is of obvious
distinction and interest.

Our results open up various questions and research avenues.  One issue is to bridge the gap between
the regime considered here and the regular thermodynamic limit.  A systematic study of overlap/work
distributions and associated quench dynamics with varying density and system sizes is currently
lacking.  A related issue is that of experimental accessibility.  Since a measurement requires
finite density, it is important to demarcate which densities in a real finite-size system show
features of our fixed-number large-size limit, from those corresponding to the regular thermodynamic
limit.  It may also be interesting to supplement our results on the Loschmidt echo with
time-evolution studies of traditionally measurable observables such as densities and correlation
functions.  Finally, Eq.\ \eqref{eq:p_O} highlights the general lack of knowledge about expectation
values of observables in (and overlaps of states with) general eigenstates of many-body
Hamiltonians.  Current activity on the Eigenstate Thermalization Hypothesis is addressing eigenstate
expectation values of some observables \cite{RigolOlshanii_Nature2008, ETH_collection}, but clearly
further investigations are warranted.

% \acknowledgments

\paragraph*{Acknowledgments.}

MH thanks M.~Vojta for discussions on equilibrium properties of the Hamiltonian \eqref{eq:minimalKondoHam}.

%%%%%%%%%%%%%%%%%%%%%%%%%%%%%%%%%%%%%%%%%%%%%%%%%%%%%%%%%%%%%%%%%%%%%%%%%%%%%%%%%%%%%%%%%%%%%%%%%%%%%%%%%%
%%%%%%%%%%%%%%%%%%%%%%%%%%%%%%%%%%%%%%%%%%%%%%%%%%%%%%%%%%%%%%%%%%%%%%%%%%%%%%%%%%%%%%%%%%%%%%%%%%%%%%%%%%

\newpage

\begin{center} \Large Supplementary Materials  \end{center}

%%%%%%%%%%%%%%%%%%%%%%%%%%%%%%%%%%%%%%%%%%%%%%%%%%%%%%%%%%%%%%%%%%%%%%%%%%%%%%%%%%%%%%%%%%%%%%%%%%%%%%%%%%
%%%%%%%%%%%%%%%%%%%%%%%%%%%%%%%%%%%%%%%%%%%%%%%%%%%%%%%%%%%%%%%%%%%%%%%%%%%%%%%%%%%%%%%%%%%%%%%%%%%%%%%%%%

%%%%%%%%%%%%%%%%%%%%%%%%%%%%%%%%%%%%%%%%%%%%%%%%%%%%%%%
% Counter resetting for supplementaries

\setcounter{figure}{0}   \renewcommand{\thefigure}{S\arabic{figure}}

\setcounter{equation}{0} \renewcommand{\theequation}{S.\arabic{equation}}

\setcounter{section}{0} \renewcommand{\thesection}{S.\Roman{section}}

\renewcommand{\thesubsection}{S.\Roman{section}.\Alph{subsection}}

% referring to subsections: don't prefix by section because that's already contained in
% \thesubsection.  Prefixes can be set using \p@*****.  Needs to be set within \makeatletter and
% \makeatother so that @ can be used in this way --->  

\makeatletter
\renewcommand*{\p@subsection}{}  
\makeatother

\renewcommand{\thesubsubsection}{S.\Roman{section}.\Alph{subsection}-\arabic{subsubsection}}

\makeatletter
\renewcommand*{\p@subsubsection}{}  % referring to subsubsections
\makeatother

%%%%%%%%%%%%%%%%%%%%%%%%%%%%%%%%%%%%%%%%%%%%%%%%%%%%%%%

%%%%%%%%%%%%%%%%%%%%%%%%%%%%%%%%%%%%%%%%%%%%%%%%%%%%%%%%%%%%%%%

\section{Overview}

In these Supplementary Materials, 

\begin{itemize}
\addtolength{\parskip}{-0.8\parskip}
\addtolength{\itemsep}{-0.8\itemsep}
\item We provide a derivation of the existence of an extended intermediate-time power-law region in the Loschmidt
  echo  $\mathcal{L}(t)$, when the work distribution $p(\omega)$ has a power-law behavior $\omega^{-b}$ with
  exponent $b\in(1,3)$.  (Section \ref{sec_suppl_intermediate_region_from_power_law}.)
\item Considering the case where the density of states has power-law behavior, we derive a relation between
  various exponents.  (Section \ref{sec_suppl_relatn_exponents}.)

\item We give some details for quantum quenches in four low-density systems.  (Interaction quench in
  Bose Hubbard chain, trap quench in Bose Hubbard chain, several quenches in hard-core bosons on a
  ladder geometry, and quasiperiodic potential quench in an Aubry-Andr\'e lattice.)  We show various
  examples of super-linear $p(\omega)\sim\omega^{-b}$ behaviors, and examples of the
  intermediate-time power-law region in $1-\mathcal{L}(t)$.
  (Section \ref{sec_supplement_various_models}.)

\item We show time evolution for the Kondo-like model with $N_c=3$ fermions, to show that there is
  no equilibration \ref{sec_suppl_3el_DE}.  (Only $N_c=1$ is shown in the main text.)

\item We provide additional details and derivations for local quenches in the Kondo-like model for
  $N_c=1$.  (Section \ref{sec_suppl_minimal_Kondo}.)

\end{itemize}

%%%%%%%%%%%%%%%%%%%%%%%%%%%%%%%%%%%%%%%%%%%%%%%%%%%%%%
\section{Derivation of the intermediate regime for $\mathcal{L}(t)$}  \label{sec_suppl_intermediate_region_from_power_law}

In this Section, we prove that an edge singularity $p(\omega)\sim\omega^{-b}$ with $b\in(1,3)$ leads
to the intermediate time behavior ${\sim}t^{b-1}$ for $\mathcal{L}(0)-\mathcal{L}(t)$, provided that
contributions of parts of  $p(\omega)$ outside this power-law region can be neglected.

We consider $p(\omega)$ to have power-law behavior $p(\omega)\sim\omega^{-b}$ in the energy window
$\omega\in[\Delta,\Lambda]$.  Here $\Delta$ is an energy scale of the order of the finite-size gap,
which should vanish in the $L\to\infty$ limit. The energy $\Lambda$ up to which the power law holds
is system-dependent, generally of the order of the bandwidth.
There are two time scales $\Lambda^{-1}$ and $\Delta^{-1}$ corresponding to these energy scales.
We show below that, when the exponent $b$ is in the appropriate range, $b\in(1,3)$, the novel
intermediate-time regime $1-\mathcal{L}(t)\sim t^{b-1}$ appears in the time window between these two
timescales, when contributions from $\omega\notin[\Delta,\Lambda]$ are neglected.

The Loschmidt echo $\mathcal{L}(t)$ is related to the Fourier transform of the work distribution:
\begin{equation}
G(t)=\int p(\omega) e^{it\omega}d\omega
\end{equation}
as $\mathcal{L}(t)=G^*(t)G(t)$.  

\begin{figure}[h]
\centering  \includegraphics[width=0.75\columnwidth]{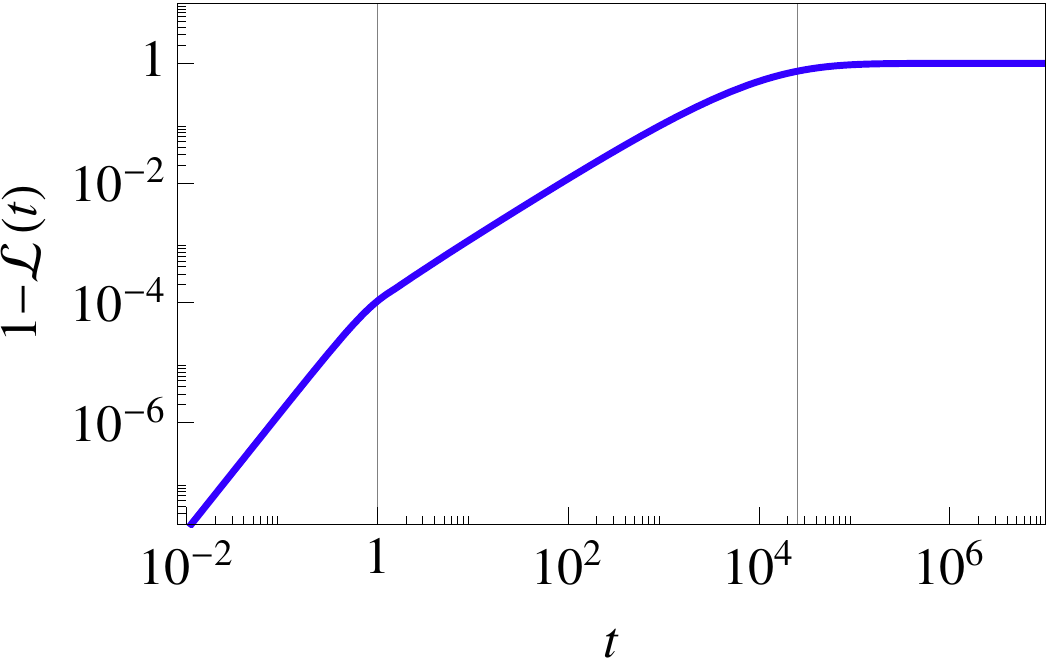}
\caption{  \label{fig_suppl_LE_analytic} 
Time evolution of the Loschmidt echo $\mathcal{L}(t)=G(t)G^*(t)$ obtained from expression
(\ref{eq:LE_exact}) with $b=2.5$, $\Delta=10^{-4}$, $\Lambda=4$.
}
\end{figure}

Neglecting the contributions from energies outside the power-law region, we obtain
\begin{multline}  \label{eq:LE_exact}
G(t) \propto \int_{\Delta}^{\Lambda}\omega^{-b}e^{it\omega}d\omega\\
~=~ \Lambda^{1-b}\mathcal{E}_{b}(-it\Lambda)-\Delta^{1-b}\mathcal{E}_{b}(-it\Delta),
\end{multline}
where $\mathcal{E}_{b}(z)=\int_1^{\infty} e^{-z\omega} \omega^{-b} d\omega$ is the generalized
exponential integral ($b$, $z\in\mathbb{C}$).  When $1-\mathcal{L}(t)$ is plotted using the above
expression \eqref{eq:LE_exact}, we see a clear intermediate-time ${\sim}t^{b-1}$ region between a ${\sim}t^2$ and a
${\sim}t^0$, as shown in Fig.\ \ref{fig_suppl_LE_analytic} for $b=2.5$.

%%%%%%%%%%% FIGURE %%%%%%%%%%%%%% FIGURE %%%%%%%%%%%%%% FIGURE %%%%%%%%%%%%%%
%%%%%%%%%%% FIGURE %%%%%%%%%%%%%% FIGURE %%%%%%%%%%%%%% FIGURE %%%%%%%%%%%%%%

Below we derive the behaviors of $\mathcal{L}(t)$ at short and intermediate times using expansions
of the exponential integral $\mathcal{E}_{b}(z)$.

\subsubsection{Short time regime.}

When $t\Lambda\ll1$, both terms in \eqref{eq:LE_exact} can be expanded in Taylor series: 
\begin{multline} \label{eq:Taylor_exp_1}
\Delta^{1-b}\mathcal{E}_{b}(-it\Delta)  =  (-i t)^{b-1}\Gamma(1-b)+\frac{\Delta^{1-b}}{b-1} \\
+i\frac{\Delta^{2-b}}{b-2}t-\frac{\Delta^{3-b}}{2(b-3)}t^{2}+O(t^{3})  \ ,
\end{multline}
and 
\begin{multline}   \label{eq:Taylor_exp_2}
\Lambda^{1-b}\mathcal{E}_{b}(-it\Lambda) = (-i t)^{b-1}\Gamma(1-b)+\frac{\Lambda^{1-b}}{b-1} \\
+i\frac{\Lambda^{2-b}}{b-2}t-\frac{\Lambda^{3-b}}{2(b-3)}t^{2}+O(t^{3}) \ . 
\end{multline}
Using \eqref{eq:Taylor_exp_1},\eqref{eq:Taylor_exp_2} in \eqref{eq:LE_exact}, and assuming
$b\in(1,3)$ and $\Lambda^{-1}\ll\Delta^{-1}$, we get
\begin{multline}
\mathcal{L}(t)=\abs{G(t)G^{*}(t)}=1+\frac{(b-1)}{(b-3)}\Delta^{b-1}\Lambda^{3-b}t^{2} \\ +\ord(t^3)
\, .
\end{multline}
Here $G(t)$ was  normalized using the criterion  $\mathcal{L}(0)=1$.

\subsubsection{Intermediate-time regime.}

In the intermediate time interval $\Lambda^{-1}\ll t\ll \Delta^{-1}$ the expression for
$\Delta^{1-b}\mathcal{E}_{b}(-it\Delta)$ is still given by the Taylor expansion
\eqref{eq:Taylor_exp_1}, but we cannot use \eqref{eq:Taylor_exp_2} because $t\Lambda$ is not small.
Instead we use the large-time asymptotics for  $\mathcal{E}_{b}(-it\Lambda)$:
\begin{equation}  \label{eq:suppl_large_tLambda}
\Lambda^{1-b}\mathcal{E}_{b}(-it\Lambda)=e^{it\Lambda}\left(\frac{i\Lambda^{-b}}{t}+O(t^{-1}\Lambda^{-1})\right) .
\end{equation}

\paragraph*{Case 1: $1<b<3$.}  

When $b$ is in this range, using the expression \eqref{eq:Taylor_exp_1} for small $t\Delta$ and
\eqref{eq:suppl_large_tLambda} for large $t\Lambda$, we obtain,
\begin{multline} 
G(t) \propto - \frac{\Delta^{1-b}}{b-1} - \Gamma(1-b)\cos\left[\frac{\pi}{2}(1-b)\right]t^{b-1}  \\
 -i\frac{\Delta^{2-b}}{b-2}t  - i\Gamma(1-b)\sin\left[\frac{\pi}{2}(1-b)\right]t^{b-1} \, .
\end{multline}
As before, $G(t)$ has to be normalized such that $\mathcal{L}(0)=1$.  Then the Loschmidt echo at leading order is
\begin{equation}
\mathcal{L}(t)=1+2(1-b)\Gamma(1-b)\cos\left[\frac{\pi}{2}(1-b)\right]\Delta^{b-1}t^{b-1}  \ .
\end{equation}

%% \textcolor{red}{Check with Yulia}

%% In the $N_c=1$ Kondo-like model (*** quenches) one obtains
%% %
%% \begin{eqnarray} \label{eq:LE_extend}
%% \mathcal{L}(t)=1+\frac{3\Gamma(-3/2)}{\sqrt{2}}\Delta^{3/2}t^{3/2}
%% \end{eqnarray} 

\paragraph*{Case 2: $b>3$.} 

In this case the leading term is  $(t\Delta)^2$ rather than $(t\Delta)^{b-1}$:
\begin{multline} 
G(t) \propto -\frac{\Delta^{1-b}}{b-1}  +\frac{\Delta^{3-b}}{2(b-3)}t^{2}
\\ - t^{b-1}\cos\left[\frac{\pi}{2}(1-b)\right]\Gamma(1-b)
-i\frac{\Delta^{2-b}}{b-2}t    \, .
\end{multline}
This gives, for the Loschmidt echo at leading order,
\begin{equation}
\mathcal{L}(t)=1-\frac{b-1}{(b-3)(b-2)^{2}}\Delta^{2}t^{2}   \, .
\end{equation}
Therefore there is no observable intermediate-time region with exponent different from 2.

%%%%%%%%%%%%%% RELATIONSHIP BETN EXPONENTS %%%%%%%%%%%%%%%%%%%%%
\section{Relationship between power-law exponents in overlap distribution, work distribution, and
  density of states} \label{sec_suppl_relatn_exponents}

In cases where the density of states has a power-law behavior, $\rho(\omega)\sim\omega^{\gamma}$,
the eigenstate index scales with the energy (relative to ground state) as 
\begin{equation}
m \sim \int_{-\infty}^{\omega}\,d\omega'\rho(\omega') ~\sim~ \omega^{\gamma+1}.
\end{equation}
Therefore for an overlap distribution with $m^{-\alpha}$ behavior we will have the dependence
\begin{equation}
\left|\langle\phi_0^{(i)}\left|\right.\phi_m^{(f)}\rangle\right|^2 ~\sim~
\omega^{-2\alpha(\gamma+1)}
\end{equation}
The definition of the work distribution $p(\omega)$ contains a factor of the density of states in
addition to this overlap squared; thus the exponent $b$ of the edge singularity
$p(\omega)\sim\omega^{-b}$ is 
\begin{equation}  \label{eq:suppl_relatn_exponents}
b = 2\gamma\alpha ~+~ 2\alpha ~-~ \gamma  \, .
\end{equation}
In the case of the $N_c=1$ Kondo-like model in the main text, we had $\gamma=-1/2$ (single-particle
density of states in one dimension), and $\alpha=$ 2 (1) for an A to A (C) quench.  This leads
to super-linear singularities in the work distribution: $b=\alpha+\tfrac{1}{2}=$ 5/2 (3/2) for the two quench regimes.  

A consequence of \eqref{eq:suppl_relatn_exponents} is that, if the exponent for the overlap happens
to be $\alpha=1/2$, the work distribution $p(\omega)$ will have linear singularity ($b=1$) for any
power-law form of the density of states.

Note that, in a generic multi-particle situation, we have no \emph{a priori} reason to expect a clean
power-law behavior of $\rho(\omega)$ at small $\omega$.  In Section \ref{sec_suppl_BH_Uquench}, we
will see a case where, even though $\alpha=1/2$, it can combine with a $\rho(\omega)$ with some more
complicated behavior to give a $p(\omega)$ with super-linear singularity ($b>1$).

%%%%%%%%%%%%%% OTHER MODELS %%%%%%%%%%%%%%%%%%%%%
\section{Various  models:\\ super-linear edge singularities\\ and\\ intermediate-time
  regime}  \label{sec_supplement_various_models}

%%%%%%%%%%% FIGURE %%%%%%%%%%%%%% FIGURE %%%%%%%%%%%%%% FIGURE %%%%%%%%%%%%%%
\begin{figure*}[bthp]
\begin{centering}
\includegraphics[width=0.85\textwidth]{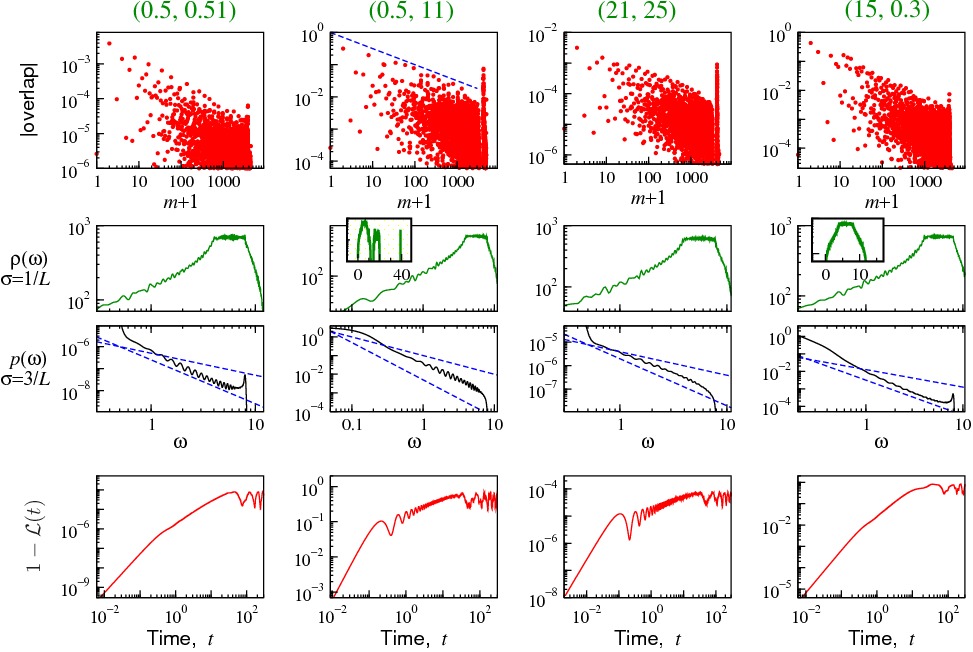}
\end{centering}
\caption{ \label{fig_suppl_BH_Uquench_3bosons} 
$N_b=3$ bosons in $L=30$ sites.  The interaction $U$ is quenched; the ($U_{\rm i}$,$U_{\rm f}$) pair
  is indicated on top of each column.  
Top row: overlap distributions.  The dashed line in second panel is ${\sim}m^{-1/2}$; the dominant
overlaps follow same exponent in every case.  For large $U_f$ (middle two panels) significant
contributions are visible from higher bands.  
Second row: many-body density of states $\rho(\omega)$ at $U=U_f$.  Insets show same with
frequencies in linear scale.  At large $U_f$, there are several bands.
Third row: Work distributions $p(\omega)$.  Dashed lines are ${\sim}b^{-3/2}$ and ${\sim}b^{-2}$.
Approximate power-law behavior of $p(\omega)$ clearly extends into $\omega$ values near the top
of the (first) band, where $\rho(\omega)$ definitely cannot be approximated with a power-law. For
large $U_f$ (middle two panels), the contributions from higher bands are not visible in the energy
range shown.   
Bottom row: Evolution of Loschmidt echo away from its initial value.  A region with intermediate
exponent between the initial ${\sim}t^2$ and the final oscillatory ${\sim}t^0$ is visible in all cases.
}
\end{figure*}
%%%%%%%%%%% FIGURE %%%%%%%%%%%%%% FIGURE %%%%%%%%%%%%%% FIGURE %%%%%%%%%%%%%%

In this section, we present numerical results on local and global quenches in several different
systems in the low-density limit.  This sampling of low-density systems shows that power-law
behaviors in the overlap distribution, and super-linear power-law behaviors in the work distribution
[$p(\omega)\sim\omega^{-b}$ with $b>1$], are generic in this important limit.  The signature in
real-time dynamics (an extended intermediate-time power-law regime), requires the additional
condition $b<3$ and that contributions outside a single power-law regime in $p(\omega)$ can be
neglected.  Therefore, the intermediate-time regime appears in some but not all cases.

In \ref{sec_suppl_BH_Uquench} we describe results for a global quench of the interaction strength,
for a Bose-Hubbard chain in the dilute limit.  In \ref{sec_suppl_BH_trapquench} we consider
Bose-Hubbard systems in a weak harmonic trap, and consider quenches of the trap strength and the
trap position.  In \ref{sec_suppl_HCB_ladder} we present results for a dilute system of hard-core
bosons in a ladder geometry, considering both local and global quenches.  Finally, in
\ref{sec_suppl_AubryAndre} we consider a single particle in a quasi-periodic (Aubry-Andr\'e)
potential, and present results for quenches of the strength of the quasi-periodic (`disorder')
potential.  For the systems of \ref{sec_suppl_HCB_ladder} and \ref{sec_suppl_AubryAndre}, we also
present some examination of the effect of increasing system size.

%%%%%%%%%%%%%%%%%%%%%%%%%%%%%%%%%%%%%%%%%%%%%%%%%%%%%%
\subsection{Bose Hubbard, interaction quench}   \label{sec_suppl_BH_Uquench}

We consider the celebrated Bose-Hubbard Hamiltonian
\begin{equation}  \label{eq_Bose_Hubbard_Hamilt}
H_{\rm BH} = -\sum_{j=1}^{L-1} b_{j}^{\dagger}b_{j+1} ~+~ \tfrac{1}{2} U \sum_{j=1}^{L}
n_{j}(n_{j}-1)
\end{equation}
in one dimension.  Here  $b_{j}$, $b_{j}^{\dagger}$ are bosonic operators for site $j$, and
$n_{j}=b_{j}^{\dagger}b_{j}$ are site occupancies.  Energies (times) are measured in units of the
hopping (inverse hopping) strength.  We use open boundary conditions and consider global quenches of
the interaction parameter $U$, form $U_i$ to $U_f$.

Figure \ref{fig_suppl_BH_Uquench_3bosons} shows numerical results for $N_{b}=3$ bosons in $L=30$
sites, which we expect to represent the ``fixed $N_b$ in large $L$'' limit.     
We show results for initial and final values of $U$ in both the small ($U\lesssim{3}$) and
large ($U\gtrsim{3}$) regimes.  The phenomena of large-exponent power-law singularity in the work
distribution and intermediate region in the time evolution of $\mathcal{L}(t)$ appear for both large
and small quenches, and for final parameter in either the large-$U$ or the small-$U$ regime.

The overlap magnitude $\left|\langle\phi_0^{(i)}\left|\right.\phi_m^{(f)}\rangle\right|$, plotted in
the top panels, naturally shows more scatter and more structures compared to the $N_c=1$ Kondo-like
model treated in the main text.  However, in each case there is a dominant set of data points which
behave as $\sim{}m^{-1/2}$.  In the large $U$ cases, where there are several `bands', there are also large
contributions from higher bands.  The extra features (the large-energy contributions, and the
contribution of the many low-energy states for which the overlap is nonzero but falls below the
dominant $m^{-1/2}$ states) will of course leave signatures in the temporal dynamics.  However, our
signature phenomenon (extended intermediate regime between perturbative and final regimes) appears in
all these cases.

In the second and third rows, we show continuous approximations for the density of states
$\rho(\omega)$ for the final $U=U_f$, and the work distribution $p(\omega)$.  This is done by replacing the delta functions
by gaussians of energy width $\sigma$, 
\begin{equation}
 \delta(\omega-\epsilon_m) ~\longrightarrow~ \frac{1}{\sqrt{2\pi}\sigma} \exp\left[
   -\frac{(\omega-\epsilon_m)^2}{2\sigma^2}\right] \, ,
\end{equation}
in the definitions
\begin{gather}
\rho(\omega) =  \sum_{m} \delta(\omega-\epsilon_m)  \\
p(\omega) = \sum_{m} \delta(\omega-\epsilon_m) \left|\langle \phi_0^{(i)}\left| \right. \phi_m^{(f)}
\rangle\right|^2  \, .
\end{gather}
Here $\epsilon_m=E_m^{(f)}-E_0^{(f)}$ are the final eigenenergies measured from the final ground
state energy.

In Figure \ref{fig_suppl_BH_Uquench_3bosons} we show continuous curves with $\sigma=L^{-1}$ for
$\rho(\omega)$ and with $\sigma=3L^{-1}$ for $p(\omega)$.  The choice of $\sigma$ is a compromise
for visualization; with smaller $\sigma$ one sees more oscillatory behavior associated with the
discreteness of the spectrum, while choosing larger $\sigma$ washes out features near the beginning
of the spectrum.

The density of states $\rho(\omega)$ has a maximum in the central region of each band.  For present
purposes, the relevant part of the spectrum is where $\rho(\omega)$ is increasing toward its first
maximum; this is naturally highlighted in the log-log plots of the second row.

In the smoothed work distributions (third row), the region $\omega\lesssim\sigma$ shows a broad
plateau which is an artifact of our smoothing procedure, and looks artificially broad on a
logarithmic scale.  We have mostly omitted this part from the plot-range.  After this part
($\omega\gtrsim\sigma$), there is an extended spectral region where $p(\omega)$ follows an
approximate power law.  The exponent $b$ of this ``edge singularity'' $p(\omega)\sim\omega^{-b}$ is
super-linear, with $b\approx2$ for small $U_f$ and $b$ slightly smaller for large $U_f$.

The approximate power-law behavior of $p(\omega)$ emerges in a more complicated manner than in the
$N_c=1$ system of the main text.  In this case, if the relevant part of the density of states
followed a power law $\rho(\omega)\sim\omega^{\gamma}$, it would not be possible to have a
super-linear edge singularity because $\alpha=1/2$ would imply $b=1$ for any $\gamma$ (see Section
\ref{sec_suppl_relatn_exponents}).  Remarkably, the non-powerlaw form of $\rho(\omega)$ conspires
with the $m^{-1/2}$ overlap distribution to cause an approximate power law in $p(\omega)$, with
exponent $b>1$.  At present it is not clear whether the power-law behavior of $p(\omega)$, or the
exponent values, becomes exact in some limit.

The $p(\omega)\sim\omega^{-b}$ behavior results in the dynamical feature that we have highlighted in
this work: in the bottom row, we see clear intermediate regimes between the perturbative and the
steady-state regimes.  The behavior in this regime is approximately $\sim{t}^\beta$, with
$\beta\approx 1$ for small $U_f$ and $\beta$ slightly smaller for large $U_f$.  This is consistent
with our prediction of $\beta=b-1$ (Section \ref{sec_suppl_intermediate_region_from_power_law}).
Note that, in the large $U_f$ cases, there are oscillations in the $\sim{t}^\beta$ region, which
result from the contribution of the higher bands.

It would be interesting to find out if an extended intermediate power-law region is visible in the
time evolution of observables that are more commonly studied (or more likely to be experimentally
measurable) than the Loschmidt echo.

%%%%%%%%%%% FIGURE %%%%%%%%%%%%%% FIGURE %%%%%%%%%%%%%% FIGURE %%%%%%%%%%%%%%
\begin{figure*}[bthp]
\begin{centering}
\includegraphics[width=0.85\textwidth]{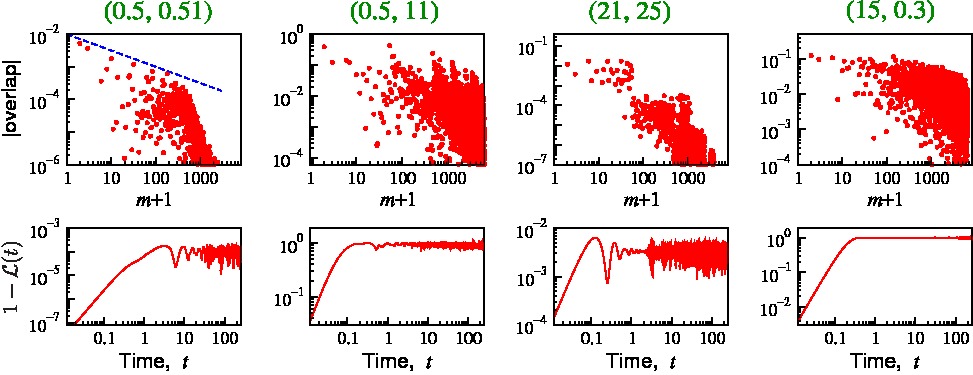}
\end{centering}
\caption{ \label{fig_suppl_BH_Uquench_unitfilling} 
$N_b=8$ bosons in $L=8$ sites.  The interaction $U$ is quenched; the ($U_{\rm i}$,$U_{\rm f}$) pair
  is indicated on top of each column.
Top row: overlap distributions.  The dashed line in first panel is ${\sim}m^{-1/2}$.  There seems to
be a power-law behavior in this weak quench, but not in any of the other quenches. 
Bottom row: Loschmidt echo.  A small  intermediate-exponent region is seen in the weak-quench
between small $U$ values, but not in any of the other quenches.   
}
\end{figure*}
%%%%%%%%%%% FIGURE %%%%%%%%%%%%%% FIGURE %%%%%%%%%%%%%% FIGURE %%%%%%%%%%%%%%

As the density (filling) is increased, the features we have presented gradually disappear.
Fig.\ \ref{fig_suppl_BH_Uquench_unitfilling} shows the case of unit filling (8 bosons in 8 sites).
The density of states and work distributions are significantly different; we do not analyze them
here.

Curiously, in weak quenches between small values of $U$ (leftmost panels), there is a sequence of
dominant overlaps that seems to follow a $m^{-1/2}$ behavior, and a small intermediate-time region
does appear in $1-\mathcal{L}(t)$.  In the corresponding work distribution (not shown), there is no
obvious $p(\omega)\sim\omega^{-b}$ behavior.  It is somewhat surprising that the intermediate-time
regime appears at unit filling.  To explain this, one could consider two different types of
$L\rightarrow\infty$ limits.  Presumably, the small intermediate-time region will disappear in the
large-$L$ limit taken with filling held constant (usual thermodynamic limit), but would get more
extended in the large-$L$ limit taken with fixed $N_b=8$.  A systematic study of the shape of the
overlap distribution, work distribution and density of states as a function of filling and size is
clearly called for, but is beyond the scope of this work.  Some $L$-dependence is explored in
Sections \ref{sec_suppl_HCB_ladder} and \ref{sec_suppl_AubryAndre}.

No intermediate-time regime is seen in the dynamics for the other quenches in
Fig.\ \ref{fig_suppl_BH_Uquench_unitfilling}; nor is there any power law like behavior in the
overlap distributions.  This is expected because the super-linear edge-singularities and the
intermediate-time regime in $\mathcal{L}(t)$ are novel features of low-density systems.

%%%%%%%%%%%%%%%%%%%%%%%%%%%%%%%%%%%%%%%%%%%%%%%%%%%%%%
\subsection{Bose Hubbard in harmonic trap; trap quench}  \label{sec_suppl_BH_trapquench}

%%%%%%%%%%% FIGURE %%%%%%%%%%%%%% FIGURE %%%%%%%%%%%%%% FIGURE %%%%%%%%%%%%%%
\begin{figure}[bthp]
\begin{centering}
\includegraphics[width=0.95\columnwidth]{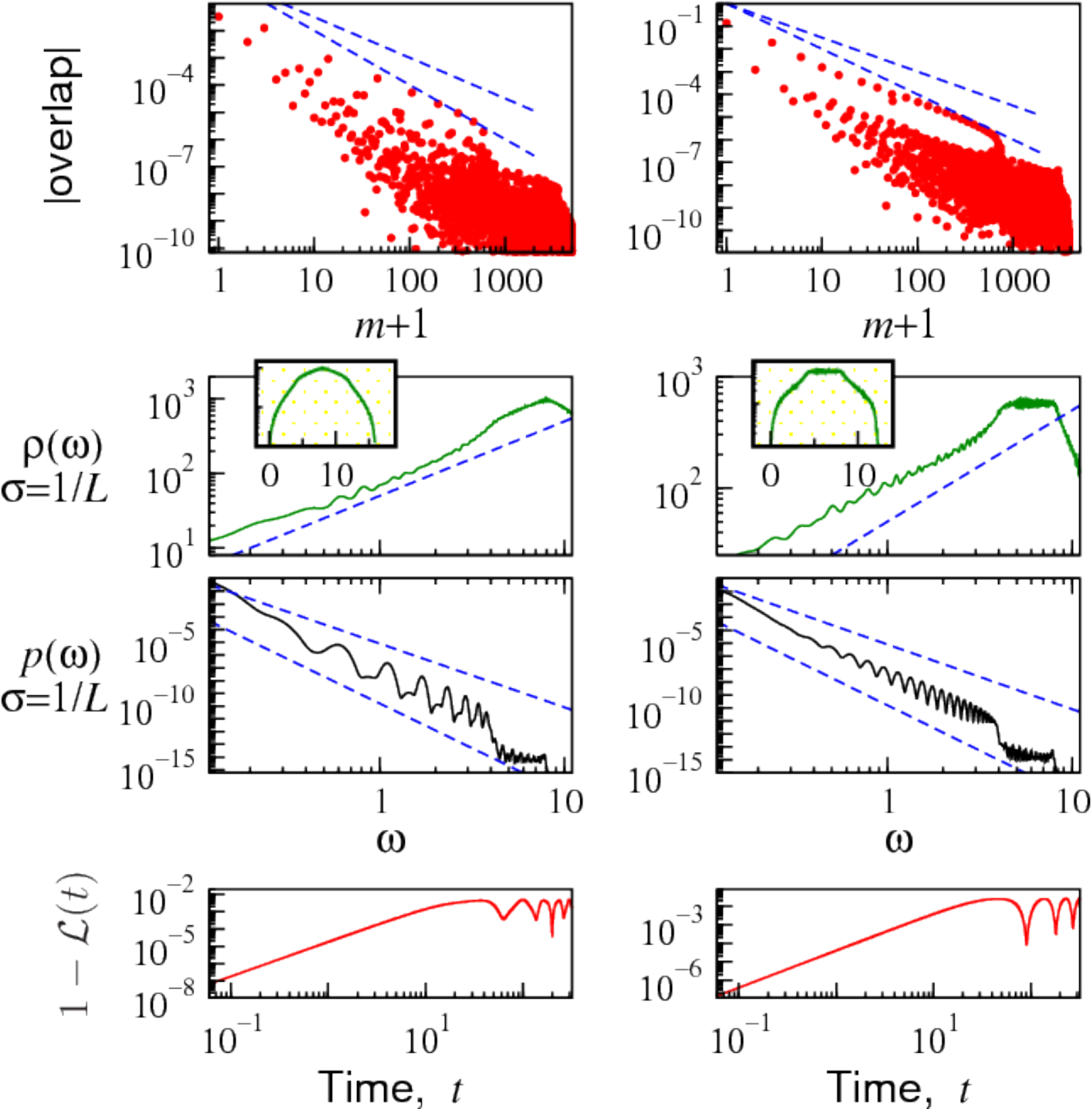}
\par\end{centering}
\caption{ \label{fig_suppl_BH_trapquench} 
Bose-Hubbard in trap:  trap strength quench (left) and trap position quench (right).
Left: $N_b=4$ bosons in $L=19$ sites; interaction $U=0.2$; trap center at $j_0=7.3$; trap strength
quenched from $k_{\rm tr}^{(i)}=0.001$ to $k_{\rm tr}^{(f)}=0.0012$.
Right: $N_b=3$, $L=28$; $U=0.2$; $k_{\rm tr}=0.002$.  Trap center quenched from $j_0^{(i)}=7.3$ to
$j_0^{(f)}=8.3$.
Dashed lines in top row (overlap panels): ${\sim}m^{-3/2}$ and ${\sim}m^{-2}$.  Dashed lines in
second row (density of states panels): $\sim\omega$.  Dashed lines in third row ($p(\omega)$
panels): ${\sim}\omega^{-5}$ and ${\sim}\omega^{-7}$.
No intermediate-exponent region is visible in the lowest row (Loschmidt echo evolution).
}
\end{figure}
%%%%%%%%%%% FIGURE %%%%%%%%%%%%%% FIGURE %%%%%%%%%%%%%% FIGURE %%%%%%%%%%%%%%

We now consider bosons on an open-boundary chain subject to to a harmonic confining trap in addition
to the Bose-Hubbard Hamiltonian \eqref{eq_Bose_Hubbard_Hamilt}:
\begin{equation}  \label{eq_trapped_Bose_Hubbard_Hamilt}
H_{\rm BH+trap} ~=~  H_{\rm BH} ~+~  \tfrac{1}{2} k_{\rm tr} \sum_{j=1}^{L} (j-j_0)^2 n_{j}
\end{equation}
Harmonic traps are fundamental to considerations of cold-atom experiments, the most prominent
experimental setting for non-equilibrium dynamics in isolated systems.  

We consider quenches of both the trap strength $k_{\rm tr}$ and the trap center $j_0$.  The
Hamiltonian \eqref{eq_trapped_Bose_Hubbard_Hamilt} has various regimes of possible interest, such as
large and small interaction $U$, strong and weak trapping potential, etc.  For present purposes, we
will confine ourselves to small $U$ and weak trapping, $k_{\rm tr}\sim\mathcal{O}(10^{-3})$.  The
weak and strong trapping regimes are loosely analogous to the A and C regimes of the Kondo-like
model detailed in the main text.

In Figure \ref{fig_suppl_BH_trapquench}, we show data for a trap strength quench ($k_{\rm
  tr}^{(i)}=0.001$ to $k_{\rm tr}^{(f)}=0.0012$) and for a trap position strength quench
($j_0^{(i)}=7.3$ to $j_0^{(f)}=8.3$).  In both cases the overlap distribution
$\left|\langle\phi_0^{(i)}\left|\right.\phi_m^{(f)}\rangle\right|$ shows a dominant series which
follows an approximate power law; the exponent $\alpha$ is between 3/2 or 2.  As in the case without
a trap (Section \ref{sec_suppl_BH_Uquench}), we show approximations to the density of states
$\rho(\omega)$ obtained by replacing delta functions with gaussians.

The density of states is similar to the case without a trap (\ref{sec_suppl_BH_Uquench}), but the
small-$\omega$ behavior of $\rho(\omega)$ seems closer to a power-law form in the presence of a
trap.  This exponent ($\gamma$) is close to 1.  The argument of \ref{sec_suppl_relatn_exponents}
then predicts the work distribution $p(\omega)\sim\omega^{-b}$ with $b=5$ (for $\alpha=3/2$) or
$b=7$ (for $\alpha=2$).  Indeed, $p(\omega)$ does have an approximate power-law decrease with
exponent around the range 5 to 7, in the energy window where $\rho(\omega)$ increases roughly
linearly.  It is difficult to be more certain about the exact values of the exponents, because of
the scatter in the overlap distribution and because of the uncertainties associated with replacing
delta functions by gaussians in order to plot $\rho(\omega)$ and $p(\omega)$.  Also, it is difficult
to be certain from the present data whether there is a genuine power-law region in $\rho(\omega)$ or
whether it is merely approximate.  Despite the difficulties with rigorous determination of
exponents, the relation \eqref{eq:suppl_relatn_exponents} is at least approximately valid.

In this low-density system and these types of quenches, we thus have edge singularities in
$p(\omega)$ with exponents far larger than that allowed in systems having he usual thermodynamic
limit.  However, as explained in Section \ref{sec_suppl_intermediate_region_from_power_law}, an
exponent in $p(\omega)\sim\omega^{-b}$ with $b>3$ does not lead to a distinguishable
intermediate-time region in the evolution of the Loschmidt echo.  Indeed the bottom panels of Figure
\ref{fig_suppl_BH_trapquench} show the ${\sim}t^2$ region directly followed by the oscillatory
region.  The singularity exponent is \emph{too large} to see an intermediate-time signature in
$\mathcal{L}(t)$.  Of course, an extended intermediate region might be visible in some observable
other than the Loschmidt echo.  An exploration of the evolution of physical observables in such
quenches remains an open task for future work.

%% A curious side note: the flat region in the density of states $\rho(\omega)$ in the middle of the
%% band is reflected as a flat plateau in the work distribution $p(\omega)$ (Figure
%% \ref{fig_suppl_BH_trapquench} right middle panels).  In Figure \ref{fig_suppl_BH_trapquench} left,
%% the density of states does not have a distinguishable flat region but a maximum in the middle of the
%% band, and this seems enough to cause a plateau in $p(\omega)$.

%%%%%%%%%%%%%%%%%%%%%%%%%%%%%%%%%%%%%%%%%%%%%%%%%%%%%%
\subsection{Hard-core bosons in a ladder}  \label{sec_suppl_HCB_ladder}

%%%%%%%%%%% FIGURE %%%%%%%%%%%%%% FIGURE %%%%%%%%%%%%%% FIGURE %%%%%%%%%%%%%%
%%%%%%%%%%% FIGURE %%%%%%%%%%%%%% FIGURE %%%%%%%%%%%%%% FIGURE %%%%%%%%%%%%%%
\begin{figure*}[tbhp]
\includegraphics[width=0.85\textwidth]{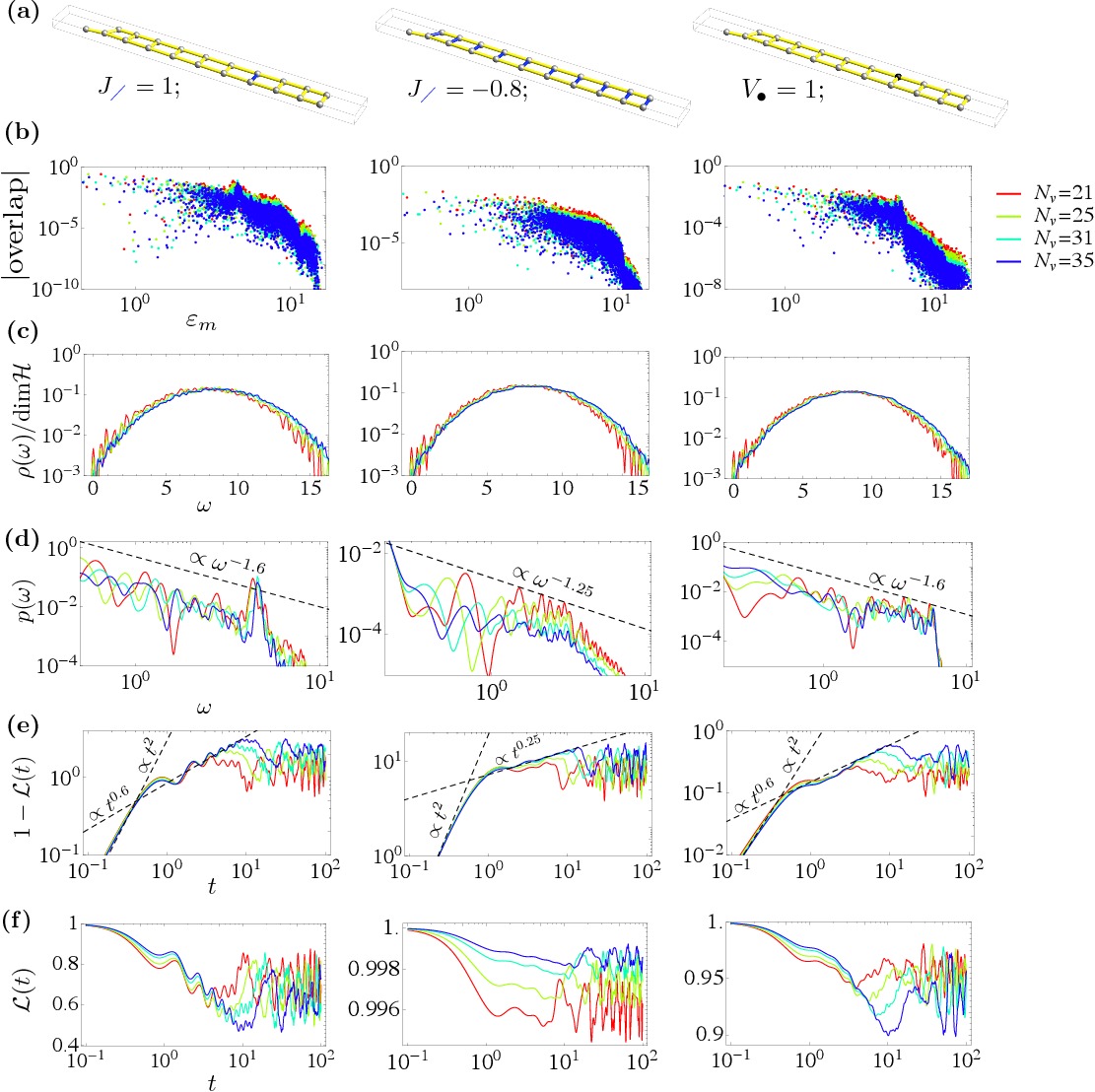}
\caption{   \label{fig:LE_HCB} 
%
% \textcolor{red}{(A) REWRITE CAPTION; (B) NEED HIGHER RESOLUTION, REPLACE FIGURE} \\ 
%
Three different quenches for a system of  $N_{b}=3$ hard-core bosons in a ladder geometry with
different ladder sizes.   
(a) Geometry of the system and quench type.  Yellow bonds correspond to $J_{i,j}=-1$ and
gray vertices to $V_{i}=0$.  Blue bonds and black vertices show parameters changed in
the quench.  
(b) Eigenstate overlap as function of energy relative to final ground state energy.   
(c) Density of states normalized to the total number of states.
(d) Work distribution.  Black dashed lines are $\sim\omega^{-b}$ with exponents $b$ chosen by eye.
(e-f) Loschmidt echo.  In (e), the perturbative ($\sim{}t^2$) and intermediate $\sim{}t^{b-1}$
regions are highlighted with dashed black power-law lines.
}
\end{figure*}
%%%%%%%%%%% FIGURE %%%%%%%%%%%%%% FIGURE %%%%%%%%%%%%%% FIGURE %%%%%%%%%%%%%%
%%%%%%%%%%% FIGURE %%%%%%%%%%%%%% FIGURE %%%%%%%%%%%%%% FIGURE %%%%%%%%%%%%%%

%%%%%%%%%%% FIGURE %%%%%%%%%%%%%% FIGURE %%%%%%%%%%%%%% FIGURE %%%%%%%%%%%%%%
%%%%%%%%%%% FIGURE %%%%%%%%%%%%%% FIGURE %%%%%%%%%%%%%% FIGURE %%%%%%%%%%%%%%
\begin{figure*}[hbtp]
\begin{centering}
\includegraphics[width=0.85\textwidth]{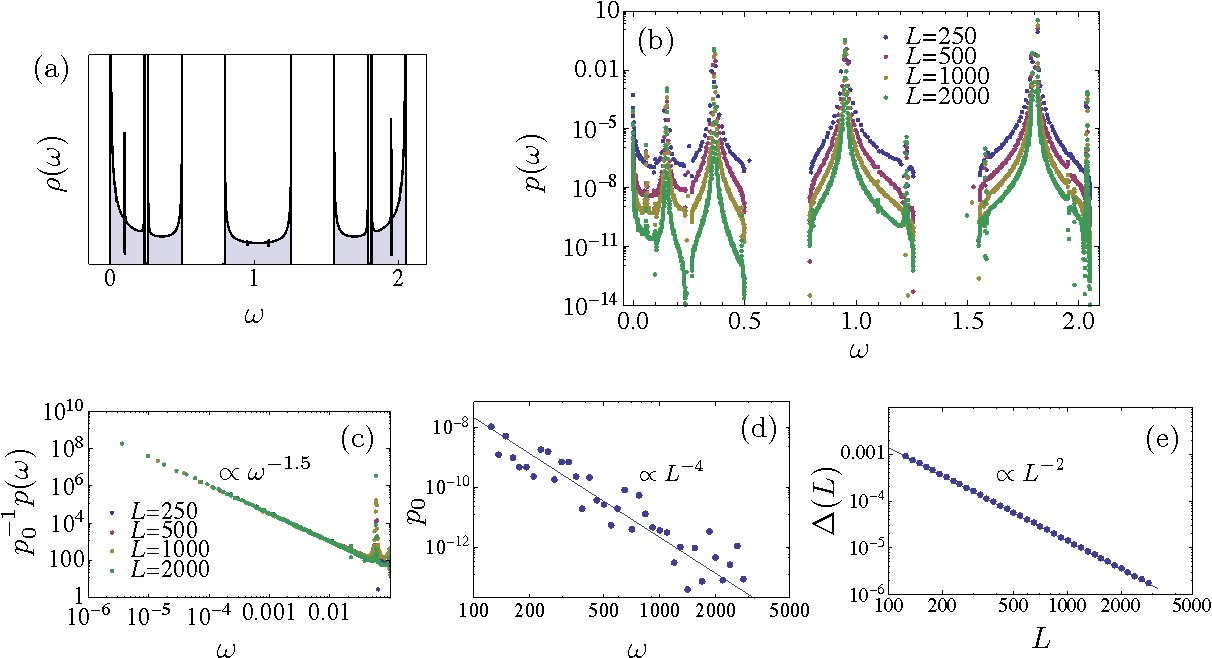}
\caption{ \label{fig_suppl_AubryAndre} 
Single particle in Aubry-Andr\'e potential.  (a) Density of states $\rho(\omega)$. (b) Work
distribution $p\left(\omega\right)$ for different system sizes $L$. 
(c) Rescaled work distributions at small $\omega$.  Values of $p_0$ are chosen such that
$p_0^{-1}p(\omega)$ for different $L$ lies on the same curve.  A power-law dependence,
$p(\omega)\sim\omega^{-b}$ with $b=3/2$, is very clear.
(d) Scaling of the pre-factor $p_{0}$ as a function of $L$.
(e) Dependence of the gap $\Delta$ on $L$.
}
\end{centering}
\end{figure*}
%%%%%%%%%%% FIGURE %%%%%%%%%%%%%% FIGURE %%%%%%%%%%%%%% FIGURE %%%%%%%%%%%%%%
%%%%%%%%%%% FIGURE %%%%%%%%%%%%%% FIGURE %%%%%%%%%%%%%% FIGURE %%%%%%%%%%%%%%

We next consider a system of $N_{b}$ hard-core bosons on a ladder of $N_{v}$ vertices with
vertex-dependent on-site potentials $V_{i}$ and bond-dependent hopping $J_{ij}$:
\begin{eqnarray}
H_{\text{HCB}} & = & \sum_{\xpct{i,j}}J_{ij}b_{i}^{\dagger}b_{j}+\sum_{i}V_{i}b_{i}^{\dagger}b_{i}\label{eq:H}
\end{eqnarray}
where $b_{i}^{\dagger}$ and $b_{i}$ are hard-core bosonic operators for site $i$ with
$\left[b_{i},b_{i}^{\dagger}\right]=1$ and $b_{i}^{\dagger2}=b_{i}^{2}=0$.  One leg of the ladder is
taken to have one site more than the other, in order to avoid spurious symmetries.

In the initial Hamiltonian, the hopping terms are set to $J_{ij}=-1$ and the on-site potentials are
set to $V_{i}=0$ . At $t=0$ some of the hopping terms or on-site potentials are changed abruptly
according to the color coding of Fig.\ \ref{fig:LE_HCB}(a) where $V_{i}=0$ is colored gray and
$J_{ij}=-1$ yellow.
We consider a local quench of a rung hopping strength ($J_{ij}$) (first column of
Fig.\ \ref{fig:LE_HCB}), a global quench of the rung hoppings (second column), and a local quench
of an on-site potential $V_i$ (third column).

Fig.\ \ref{fig:LE_HCB}(b) shows the overlaps $\abs{\braket{\phi_{0}^{(i)}}{\phi_{m}^{(f)}}}$ as a
function of the eigenenergies $\varepsilon_{m}$.  The overlaps have a maximum magnitude at $m=0$ for
all the quenches considered. The density of states $\rho(\omega)$ of the final Hamiltonian, given in
Fig.\ \ref{fig:LE_HCB}(c), was obtained by Gaussian regularization of the delta function (as in
Sections \ref{sec_suppl_BH_Uquench} and \ref{sec_suppl_BH_trapquench}) with width $\sigma=0.1$.  For
the system sizes considered here, $\rho(\omega)$ is noisy near the band edges as there are not
enough states in this regions to obtain a smooth curve. The same phenomenon appears in the work
distribution $p\left(\omega\right)$ shown in Fig.\ \ref{fig:LE_HCB}(d).  This quantity is sharply
peaked around the bottom of the spectrum.
%
%% It is difficult to assign an algebraic dependence to $p\left(\omega\right)$ with the value
%% $\sigma=0.1$ used here for regularization.  
%
The power-law lines in Fig.\ \ref{fig:LE_HCB}(d) are only guides to the eye.

Even though it is difficult to assign unambiguous power-law exponents for $p(\omega)$, the existence
of an intermediate-time regime in the temporal dynamics is clear, as seen in
Fig.\ \ref{fig:LE_HCB}(e) for $1-\mathcal{L}\left(t\right)$.  The exponents in
$p(\omega)\sim\omega^{-b}$, Fig.\ \ref{fig:LE_HCB}(d), are consistent with the exponents for the
intermediate-region ($\sim{}t^{\beta}$), Fig.\ \ref{fig:LE_HCB}(e), following the results of Section
\ref{sec_suppl_intermediate_region_from_power_law}: $\beta=b-1$.

Fig.\ \ref{fig:LE_HCB}(e) also shows the perturbative region for small $t$ and the oscillatory
behavior obtained at large times.  The time at which the large-time oscillatory behavior starts
(corresponding to $\sim\Delta^{-1}$ of Section \ref{sec_suppl_intermediate_region_from_power_law})
clearly grows with system size.  
The extent of the intermediate-time regime gets progressively wider as the system size increases at
fixed particle number.

Finally Fig.\ \ref{fig:LE_HCB}(g), shows how $\mathcal{L}(t)$ itself (not subtracted from
$\mathcal{L}(0)=1$) looks like in the intermediate-time region.

%%%%%%%%%%%%%%%%%%%%%%%%%%%%%%%%%%%%%%%%%%%%%%%%%%%%%%
\subsection{Single particle in Aubry-Andr\'e potential}  \label{sec_suppl_AubryAndre}

We consider a single particle on an $L$-site chain subject to a quasiperiodic potential, i.e., a
cosine potential with incommensurate period:
\begin{equation}
H  =  -J\sum_{j=0}^{L-2}\left( c_j^{\dagger}c_{j+1} +  c_{j+1}^{\dagger}c_{j}
\right)\label{eq:hamiltonian-definition}   ~+~ \sum_{j} V_j c_j^{\dagger}c_{j}
\end{equation}
with $V_j =V\cos\left(2\pi q_{1}j\right)$ having irrational wave vector $q_{1}$, here taken to be
$q_{1}=\frac{\sqrt{5}-1}{2}$.  This is known as the Aubry-Andr\'e model.  There is a localization
transition at $V/J=2$; single-particle eigenstates are exponentially localized in space for $V/J>2$.
We will consider quenches within the delocalized regime, from $V=0.1J$ to $V=0.6J$.

As $V$ is changed from 0 toward $2J$, the tight-binding band splits up into sub-bands and continues
splitting further until it becomes fractal at $V=2J$.  Fig.\ \ref{fig_suppl_AubryAndre}(a) shows the
density of states for $V=0.6J$, where the (sub)band edges are seen through $|\omega-\omega_{\rm
  \small edge}|^{-1/2}$ cusps characteristic of single-particle bands in 1D.  (The energy gaps
between (sub)bands are sometimes too small to be seen.)  Fig.\ \ref{fig_suppl_AubryAndre}(b) shows
the work distribution $p(\omega)$.  The work distribution shows edge singularities at every band
edge and is thus quite intricate.  Fig.\ref{fig_suppl_AubryAndre}(c) shows that the work
distribution near the bottom of the first band ($\omega{\to}0$) has a clear power law behavior;
$p\left(\omega\right)\propto\omega^{-3/2}$.  As expected, in order to fulfill the normalization
condition $\int d\omega p\left(\omega\right)=1$, the magnitude of $p(\omega)$ decreases with $L$.
The pre-factor $p_0$ decreases as $L^{-4}$ overall [Fig.\ \ref{fig_suppl_AubryAndre}(d)].  There are
some fluctuations because of the quasi-random nature of the system.  The gap between the ground and
the first excited state, $\Delta$, shown in Fig.\ \ref{fig_suppl_AubryAndre}(e), shows a
characteristics $L^{-2}$ decay with $L$.

Because $p(\omega)$ has large contributions in several regions outside the small-$\omega$ power-law
region, no intermediate-time region is visible in the time evolution of the Loschmidt echo.

%% Summarizing our findings, in the $\omega\to0$ limit one has  
%% \begin{eqnarray*}
%% p\left(\omega\right) & = & \Theta\left[\omega-\Delta_{0}\left(L\right)\right]\frac{L^{-\beta}}{\Delta_{0}\left(L\right)}\left|\frac{\omega}{\Delta_{0}\left(L\right)}\right|^{-b}
%% \end{eqnarray*}
%% with 
%% \begin{eqnarray*}
%% b & = & 1.5\\
%% \beta & = & 3\\
%% \Delta_{0}\left(L\right) & = & \delta_{0}L^{-2}
%% \end{eqnarray*}

%% Note that the weight of the low energy part $\int_{\Delta}^{\Lambda}d\omega\, p\left(\omega\right)\propto L^{-\beta}$
%% with $\Lambda\simeq0.01$, vanishes with $L$. Because of the peak
%% like structure of $p(\omega)$ and the fact that the weight of the
%% low energy part vanishes with $L$, the extended region is not observed
%% in the $\mathcal{L}\left(t\right)$. 

%%%%%%%%%%%%%%%%%%%%%%%%%%%%%%%%%%%%%%%%%%%%%%%%%%%%%%
\section{Lack of equilibration in Kondo-like model with $N_c=3$ fermions}  \label{sec_suppl_3el_DE}

\begin{figure}[btp]
\includegraphics[width=0.85\columnwidth]{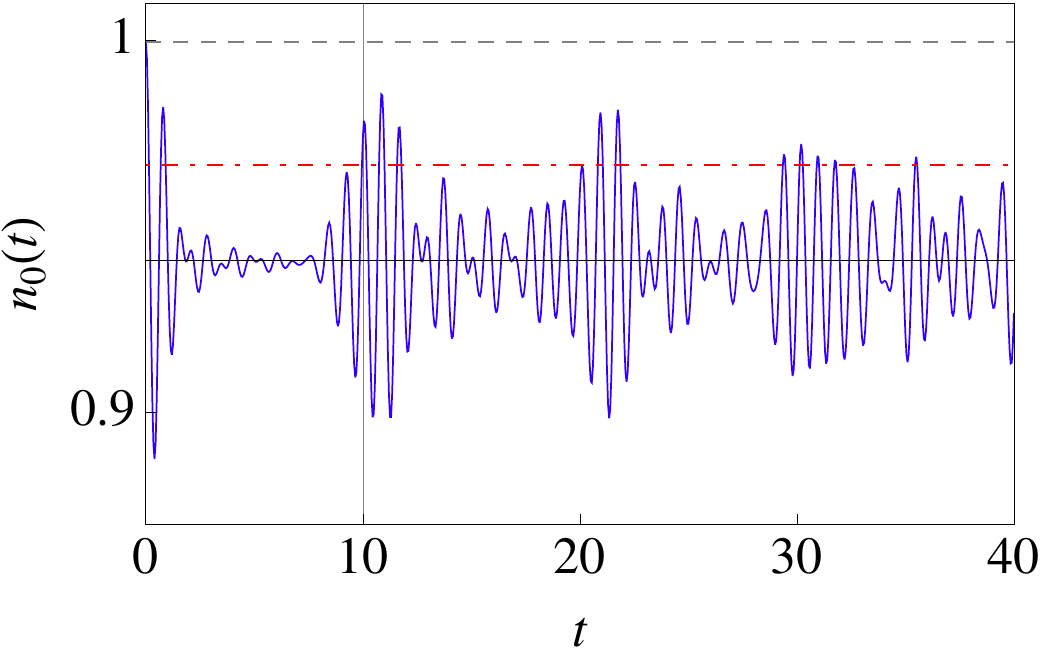}
\caption{ \label{fig:DE3el} 
Blue solid curve: Time evolution of the local density $n_0(t)$ in the Kondo-like model,
with $N_c=3$ fermions, $L=20$, $J_i=10$, $J_f=100$. 
Solid black line: the long-time average value of  $n_0(t)$, i.e., the diagonal average value
$\xpct{n_0}_{DE}$.  
The gray dashed and red dot-dashed lines indicate the ground-state values of $n_0$ corresponding to
$J=J_i$ and to $J=J_f$.
}
\end{figure}

In the main text we presented the absence of equilibration to the final equilibrium value for
$N_c=1$ fermion, in contrast to the expectation for local quenches in finite-density systems.
This is not a single-particle curiosity, but a generic feature of low-density systems.  In
Fig.\ \ref{fig:DE3el} we show an example of the time evolution of the local occupancy at the
impurity-coupled site, $n_0(t)$, after a quench from $J_i=10$ to $J_f=100$.  The long-time average
$\xpct{n_0}_{\mathrm{DE}}$ is the black solid horizontal line around which $n_0(t)$ oscillates.  This is
markedly different from the final equilibrium value, shown as the red dot-dashed horizontal line.

%%%%%%%%%%% FIGURE %%%%%%%%%%%%%% FIGURE %%%%%%%%%%%%%% FIGURE %%%%%%%%%%%%%%
%%%%%%%%%%% FIGURE %%%%%%%%%%%%%% FIGURE %%%%%%%%%%%%%% FIGURE %%%%%%%%%%%%%%

%%%%%%%%%%%%%%%%%%%%%%%%%%%%%%%%%%%%%%%%%%%%%%%%%%%%%%%%%%%%%%%
\section{Details for $N_c=1$ Kondo-like model}  \label{sec_suppl_minimal_Kondo}

In this Section we provide details for results presented in the main text for the Kondo-like model
with a single mobile fermion ($N_c=1$).
In \ref{sec_suppl_overlap_behaviors} we provide derivations of the forms of the overlap
distributions presented in Figure 3 of the main text.  
In \ref{sec_suppl_A2A_powerlaw_scaling} we show how parameter values for A$\to$A quenches can be
chosen in a way such that power law region of the overlap distribution gets systematically extended with system
size.
In \ref{sec_suppl_A2C_noIntermediate} we explain why for A$\to$C quenches there is no
intermediate-time region visible in $n_0(t)$, even though such a region is visible in
$\mathcal{L}(t)$, as seen in Figure 4 of the main text.

\paragraph*{Spatial index as basis label.} 

Since we are restricted to the singlet sector, for $N_c=1$, we can use the position of the mobile
fermion as the label for a complete set of states spanning the singlet sector:
\begin{equation}
\ket{j} = \frac{1}{\sqrt{2}} \left( c_{j\downarrow}^{\dagger}\ket{\Omega} \ket{\uparrow} -
c_{j\uparrow}^{\dagger}\ket{\Omega} \ket{\downarrow} \right)
\end{equation}
where $j$ is the site index and $\ket{\Omega}$ is the fermionic vacuum.

%%%%%%%%%%%%%%%%%%%%%%%%%%%%%%%%%%%%%%%%%%%%%%%%%%%%%%
\subsection{Derivation of the overlap behaviors} \label{sec_suppl_overlap_behaviors} 

We summarize below analytic calculations for the overlaps
$\abs{\braket{\phi_{0}^{(i)}}{\phi_{m}^{(f)}}}$ for quench cases starting and ending in A and C
regimes.

We start with A$\to$A quenches, $J_{i,f} \ll 1$.  We derive the overlap behavior (power law,
${\sim}m^{-2}$) by treating $J_{i,f}$ perturbatively.  Next, we derive the overlap behaviors for
quenches between A and C regimes when $J_{i(f)}{\ll}1$ and $J_{f(i)}{\gg}1$, by using zeroth-order
expressions for the eigenfunctions in the two limits.  For C$\to$C quenches, we derive the sine
behavior of the overlap by using the zeroth-order ($J{\gg}1$) expression for the final eigenstates.

In the main text, the index $m$ was used to label the eigenstates within the symmetry sector where
the z-component of the total spin is zero.  Within this symmetry sector, there are also triplet
states and states with odd spatial reflection parity, which have zero overlap with the initial
state.  Here, we restrict further to eigenstates with nonzero overlap, which are about one-fourth of
the states spanned by the $m$ index.  We use $\mu$ to index this restricted set of eigenstates.

%%%%%%%%%%% FIGURE %%%%%%%%%%%%%% FIGURE %%%%%%%%%%%%%% FIGURE %%%%%%%%%%%%%%
\begin{figure}[tbp]
\centering
\includegraphics[width=0.95\columnwidth]{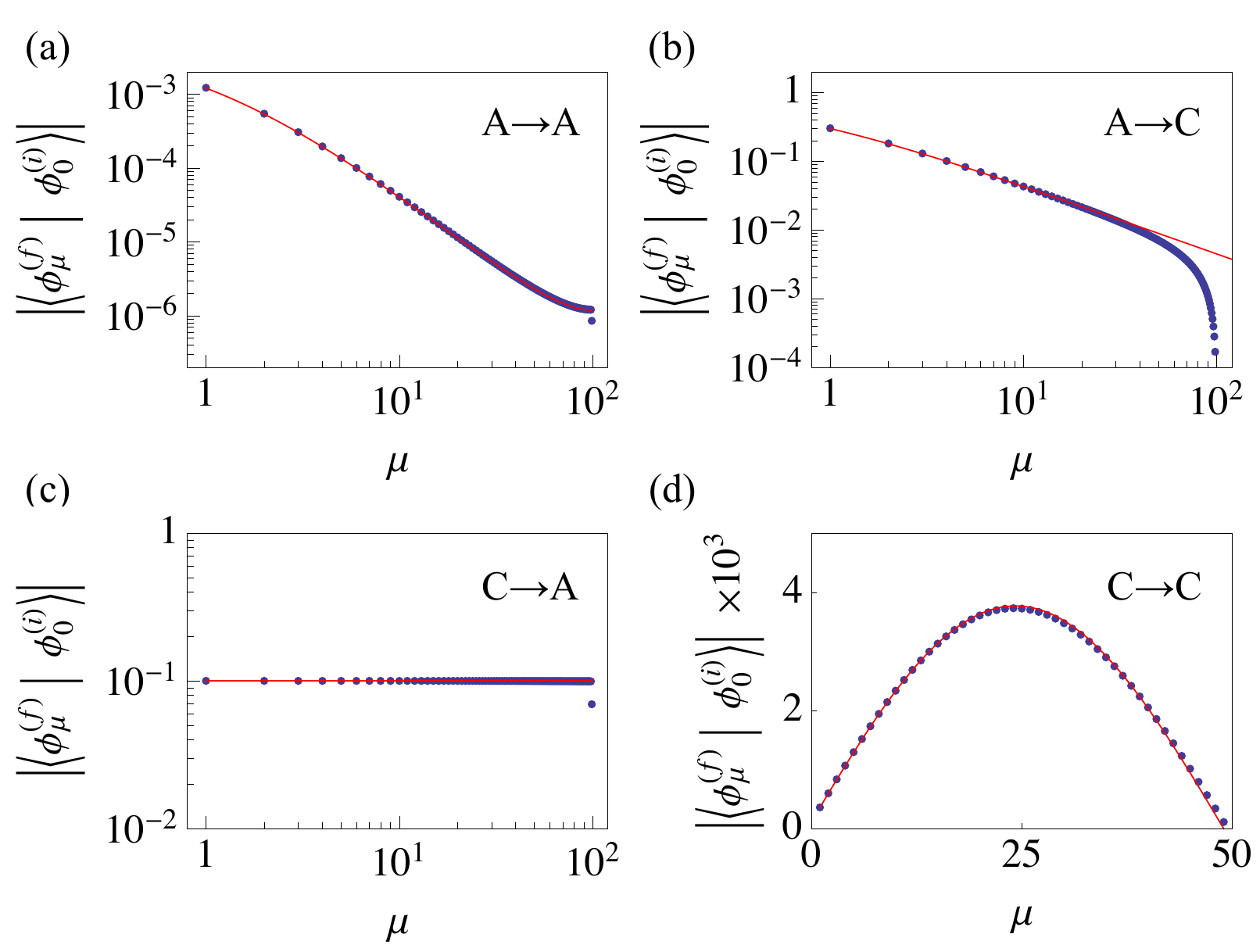}
\caption{ \label{fig_suppl_1el_overlaps_derivations}
Overlap distributions for the $N_c=1$ Kondo-like model with $L=100$, compared with analytic
predictions.  Only the even-parity excited singlet states are indexed by $\mu$.
(a) A$\to$A quench, $J_{i}=10^{-3}$, $J_{f}=10^{-4}$, Red solid line is Eq.\ \eqref{eq:overlap_AA}.
(b) A$\to$C quench, $J_{i}=10^{-3}$, $J_{f}=10^{3}$.  Red solid line is Eq.\ \eqref{eq:overlap_AC}.
(c) C$\to$A quench, $J_{i}=10^{3}$, $J_{f}=10^{-3}$.  Red solid line is Eq.\ \eqref{eq:overlap_CA}.
(d) C$\to$C quench, $J_{i}=10^{2}$, $J_{f}=10^{4}$.  Red solid line is Eq.\ \eqref{eq:overlap_CC}.
}
\end{figure}
%%%%%%%%%%% FIGURE %%%%%%%%%%%%%% FIGURE %%%%%%%%%%%%%% FIGURE %%%%%%%%%%%%%%

\subsubsection{A$\to$A quench}  

For  $J_{i,f}{\ll}1$, we write both initial and final Hamiltonians as  $\hat{H}_{i,f}= \hat{H}_0 +
J_{i,f}\hat{V}$ and use first-order perturbation theory.  

The reflection-symmetric eigenfunctions of the unperturbed Hamiltonian $\hat{H}_0$ are the plane
waves:
\begin{equation} \label{eq:suppl_A_0th_order}
\ket{\phi_{\mu}^{A,(0)}}=\sum_{i=0}^{L-1}\sqrt{2L^{-1}}\cos\left[\frac{\pi}{L}2{\mu}j\right]\ket{j}
\end{equation}
and the energies are $E^{(0)}_{\mu}=-2\cos\left[\frac{2\pi}{L}\mu\right]$.  At first order, the
eigenfunctions are
\begin{equation}
\ket{\phi_{\mu}^{\text{A,(1)}}}=\ket{\phi_{\mu}^{\text{A,(0)}}}+\sum_{k=0, k\neq \mu}^{L/2}
\frac{\bra{\phi_{k}^{\text{A,(0)}}}J\hat{V}\ket{\phi_{\mu}^{\text{A,(0)}}}}{E_{\mu}^{(0)}-E_{k}^{(0)}}\ket{\phi_{k}^{\text{A,(0)}}}.
\end{equation}
Therefore, the overlaps of the ground state $\ket{\phi^{(i)}_0}$ of $H_{i}$ with the excited states
$\ket{\phi^{(f)}_{\mu}}$ of $H_{f}$ are, at first order,
\begin{multline}
\braket{\phi^{(i)}_0}{\phi^{(f)}_{\mu}}
~=~ \braket{\phi^{(0)}_0}{\phi^{(1)}_{\mu}} ~+~ \braket{\phi^{(1)}_0}{\phi^{(0)}_{\mu}}
\\ =~ \frac{J_f-J_i}{E^{(0)}_{\mu}-E^{(0)}_0}V^{(0)}_{\mu,0}
\, .
\end{multline}
The matrix elements $V_{\mu,0}=\bra{\phi^{(0)}_{\mu}}\hat{V}\ket{\phi^{(0)}_0}$ are found to be constant,
$V_{\mu,0}=3\sqrt{2}\left(4L\right)^{-1}$.  Thus the overlaps as function of $\mu$ are
\begin{equation}\label{eq:overlap_AA}
\abs{\braket{\phi_{0}^{A,(1)}}{\phi_{\mu}^{A,(1)}}} ~=~
\frac{3\sqrt{2}}{8L}\frac{\abs{J_f-J_i}}{1-\cos[2\pi L^{-1} \mu]} \, .
\end{equation}
This expression is compared to exact numerical values in
Fig.\ \ref{fig_suppl_1el_overlaps_derivations}(a).

\subsubsection{A$\to$C quench}

The final ground state is not included in the power-law regime for A to C quenches, reflecting the
fact that the ground state in the C regime is well-separated and quite different from the other
eigenstates.  In the calculations below, we are therefore only interested in $\mu>0$.  
For very large $J$ (extreme C regime), the particle is localized at site $i=0$ in the ground state,
and excluded from this site in the other eigenstates.  Thus the excited eigenstates (``band
states'') are the single-particle wave functions in a chain of length $L-1$ with hard-wall
boundary conditions:
\begin{equation}  \label{eq:suppl_C_0th_order}
\ket{\phi_{\mu\neq0}^{C,(0)}}=\sum_{i=1}^{L-1}\sqrt{2L^{-1}}\sin\left[\frac{\pi}{L}\left(2\mu+1\right)j\right]\ket{j}
\ .
\end{equation}
The initial ($J\ll1$) state is $\ket{\phi_{0}^{A,(0)}(0)}=L^{-1/2}\sum_{x}\ket{x}$ at lowest order.  
Therefore the overlaps are
\begin{equation}
\braket{\phi_{0}^{A,(0)}}{\phi_{\mu\neq0}^{C,(0)}} =
\frac{\sqrt{2}}{L}\sum_{x=1}^{L-1}\sin\left[\frac{2\pi}{L}\left(\mu+\frac{1}{2}\right)x\right] \, .
\end{equation}
In the limit $L\rightarrow \infty$ one can replace summation with integration, leading to 
\begin{eqnarray} \label{eq:overlap_AC}
\braket{\phi_{0}^{A,(0)}}{\phi_{\mu\neq0}^{C,(0)}} =\frac{2\sqrt{2}}{\pi(2\mu+1)}.
\end{eqnarray}

This expression is compared to exact numerical values in
Fig.\ \ref{fig_suppl_1el_overlaps_derivations}(b).

\subsubsection{C$\to$A quench}

For $J_i\gg{1}$ and $J_f\ll{1}$, we use the zeroth-order expressions in the two limits:
$\ket{\phi_0^{(i)}} = \ket{\phi_{0}^{C,(0)}}=\sum_{i=0}^{L-1}\delta_{i,0}\ket{x}$ and
$\ket{\phi_{\mu}^{(f)}} = \ket{\phi_{\mu}^{A,(0)}}$ given by Eq.\ \eqref{eq:suppl_A_0th_order}.  The
resulting overlap distribution is the constant function
\begin{equation}\label{eq:overlap_CA}
\braket{\phi_{0}^{C,(0)}}{\phi_{\mu}^{A,(0)}}=\sqrt{\frac{2}{L}}.
\end{equation}
This expression is compared to exact numerical values in
Fig.\ \ref{fig_suppl_1el_overlaps_derivations}(c).

\subsubsection{C$\to$C quench}

We prove the sine behavior for quenches from anywhere in B or C regimes to the extreme C case
($J_f=\infty$).  Describing the sine behavior more generally for anywhere in C to anywhere in C
involves the same physical ideas, but is too clumsy in notation to write out.

For the final excited states we use the zeroth order expression \eqref{eq:suppl_C_0th_order}.  For
the initial ground state, we can formally use the exact form $\ket{\phi_{0}^{(i)}} =
\mathcal{N}^{-1/2}\exp\left[ -[x]/\xi(J_i)\right]$, where $[x]=\min(x,L-x)$ is the distance from the
impurity coupled site.  Here $\xi(J)$ is a decreasing function of $J$ and
$\mathcal{N}(J)$ is a normalization factor.
This gives for the overlaps 
\begin{equation}
\braket{\phi_{0}^{(i)}}{\phi_{\mu\neq0}^{(f)}}=\frac{2\sqrt{2}}{\sqrt{L\mathcal{N}(J_i)}} \sum_{x=1}^{L/2}
\sin\left[\frac{\pi}{L}\left(2\mu+1\right)x\right] e^{-\abs{x}/\xi(J_i)} .
\end{equation}
By definition of the C regime, $\xi(J_i)<1$ and  $\exp\left[\frac{-\abs{x}}{\xi(J_i)}\right]$ is a fast
decaying function.  Hence the main contribution to the sum is given by the first term ($x=1$): 
\begin{equation}\label{eq:overlap_CC}
\braket{\phi_{0}^{C}}{\phi_{\mu}^{C,(0)}}=\frac{2\sqrt{2}}{\sqrt{L\mathcal{N}(J_i)}}
\sin\left[\frac{\pi}{L}\left(2\mu+1\right)\right] e^{-1/\xi(J_i)} . 
\end{equation}
This expression is compared to exact numerical values in
Fig.\ \ref{fig_suppl_1el_overlaps_derivations}(b).

%%%%%%%%%%%%%%%%%%%%%%%%%%%%%%%%%%%%%%%%%%%%%%%%%%%%%%%%%%%%%%
\subsection{Scaling demonstration of the power law in A to A
  quench}  \label{sec_suppl_A2A_powerlaw_scaling}

%%%%%%%%%%% FIGURE %%%%%%%%%%%%%% FIGURE %%%%%%%%%%%%%% FIGURE %%%%%%%%%%%%%%
\begin{figure}[htb]
\centering
\includegraphics[width=1\columnwidth]{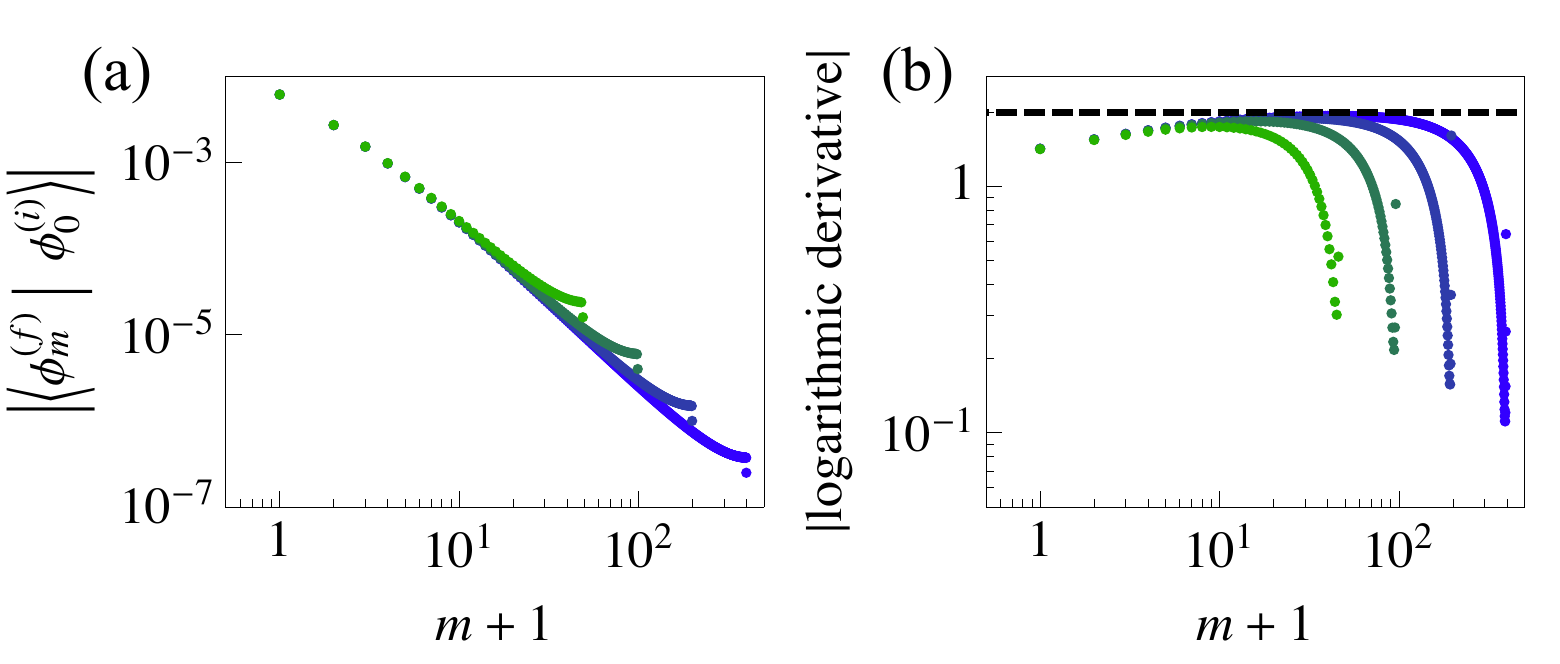}
\caption{  \label{fig_suppl_scaling}
The overlap distribution $\abs{\braket{\phi_0^{(i)}}{\phi_m^{(f)}}}$ and the numerically determined
power-law exponent (absolute value of the logarithmic derivative), plotted for different sizes.
Sizes $L=100,\:200,\:400,\:800$ are used with the $J_{i,f}$ adjusted so as to keep $\xi(J_{i,f})/L$
fixed.  For $L=100$ we used $J_{i}=10^{-3}$, $J_{f}=10^{-2}$.  
}
\end{figure}
%%%%%%%%%%% FIGURE %%%%%%%%%%%%%% FIGURE %%%%%%%%%%%%%% FIGURE %%%%%%%%%%%%%%

The A$\to$A quench overlap data in Figure 3 of the main text clearly shows a power-law behavior
$\abs{\braket{\phi_0^{(i)}}{\phi_m^{(f)}}}{\sim}m^{-2}$, for a particular system size $L$.  However,
since the definition of the $A$ region is itself $L$-dependent, the extension of this behavior with
increasing $L$ is not immediately obvious.   

In Figure \ref{fig_suppl_scaling}, we plot  A$\to$A  overlap distributions for several system
sizes.  As $L$ is varied, we adjust $J_{i,f}$ such that the ratio of  $\xi(J)$ to $L$ stays fixed.
Here $\xi(J)$ is the localization length that the mobile fermion would have in case of an infinite
system.  In the A regime, $\xi(J)>L$.  Adjusting  $J_{i,f}$ in this manner ensures that we do not
get into the B regime as $L$ is increased.  In addition, it turns out that increasing $L$ and
decreasing $J_{i,f}$ in this coordinated manner yields a series of curves that systematically
extends the power-law behavior, as seen in  Figure \ref{fig_suppl_scaling}(a).  

Figure \ref{fig_suppl_scaling}(b) shows the logarithmic derivative of the
$\abs{\braket{\phi_0^{(i)}}{\phi_m^{(f)}}}$ versus $m$ data, which extracts the power-law exponent.
We note that the region where this quantity is numerically $\alpha=2$ (dashed horizontal line), also
gets systematically extended through this procedure.

%%%%%%%%%%%%%%%%%%%%%%%%%%%%%%%%%%%%%%%%%%%%%%%%%%%%%%
\subsection{Absence of intermediate-time regime in $n_j(t)$ for A to C quench}  \label{sec_suppl_A2C_noIntermediate}

For A$\to$C quenches, an extended intermediate-exponent regime in the time evolution is seen in the
Loschmidt echo $\mathcal{L}(t)$ but not in the local density $n_0(t)$.  To clarify this situation,
we note that $\hat{n}_j$ is a rank-1 projection operator: $\hat{n}_j=\ket{j}\bra{j}$.  Using the
language introduced in the main text for such observables, instead of the work distribution we
should consider
\begin{equation} 
p_{\hat{n}_j}(\omega)=\sum_m \delta(\omega-E_m) \braket{\phi_0^{(i)}}{\phi_m^{(f)}}
\braket{\phi_m^{(f)}}{j} 
\end{equation}
One factor of the overlap distribution is now replaced by $\braket{\phi_m^{(f)}}{j}$.  In the C
region, for the excited (`band') eigenstates $m>0$, this quantity is approximately constant except
when it vanishes for symmetry reasons.  In the excited eigenstates $m>0$, the quantity is very small
for the impurity-coupled site $j=0$, and $\ord(1/L)$ for the other ($j>0$) sites.  In either case,
this leads to the power-law behavior $p_{\hat{n}_j}(\omega)\propto \omega^{-b_{\hat{n}_j}}$ with
$b_{\hat{n}_j}=1$.  This exponent is not large enough to cause a distinct intermediate-time region
in $\left|{n_j(t)-n_j(0)}\right|$.  

In contrast, the exponent relevant for the Loschmidt echo is $b=3/2$, appearing in the work
distribution.  This leads to an intermediate-exponent region $1-\mathcal{L}(t)\sim t^{1/2}$ between
the initial ${\sim}t^2$ and the large-time oscillatory ${\sim}t^0$ regimes.  (Figure 4 in main
text.)

Note that, in addition to having an unsuitable exponent, the quantity $p_{\hat{n}_0}(\omega)$
corresponding to site $j=0$ has large contribution from $m=0$, which is outside the power-law
region.  The corresponding quantities for other sites, say $p_{\hat{n}_{L/2}}(\omega)$ for $j=L/2$,
would have larger spectral weight in the power-law region.  However, since $b_{\hat{n}_j}=1$ for all
sites $j$, there is still no intermediate-time region for $j\neq{0}$.

\end{document}